# Impact of electron beam propagation on high-resolution quantitative chemical analysis of 1-nm-wide GaN/AlGaN quantum wells

Florian Castioni[1,2,*], Patrick Quéméré[1], Sergi Cuesta[3], Vincent Delaye[1], Pascale Bayle-Guillemaud[3], Eva Monroy[3], Eric Robin[3], Nicolas Bernier[1]

[1] Univ. Grenoble Alpes, CEA, LETI, 38000 Grenoble, France

[2] Univ. Paris-Saclay, CNRS, Laboratoire de Physique des Solides, 91405, Orsay, France

[3] Univ. Grenoble Alpes, CEA, IRIG, 38000 Grenoble, France

* Corresponding author: florian.castioni@universite-paris-saclay.fr

**Abstract:**

Recent advancements in high-resolution spectroscopy analyses within the scanning transmission electron microscope (STEM) have paved the way for measuring the concentration of chemical species in crystalline materials at the atomic scale. However, several artifacts complicate the direct interpretation of experimental data. For instance, in the case of energy dispersive x-ray (EDX) spectroscopy, the linear dependency of local x-ray emission on composition is disrupted by channeling effects and cross-talk during electron beam propagation. To address these challenges, it becomes necessary to adopt an approach that combines experimental data with inelastic scattering simulations. This method aims to account for the effects of electron beam propagation on x-ray emission, essentially determining the quantity and the spatial origin of the collected signal. In this publication, we propose to assess the precision and sensitivity limits of this approach in a practical case study involving a focused ion beam (FIB)-prepared III-N multilayers device. The device features nominally pure ~1.5-nm-wide GaN quantum wells surrounded by AlGaN barriers containing a low concentration of aluminum (~5 at. %). By employing atomic-scale EDX acquisitions based on the averaging of more than several thousand frames, calibrated ζ-factors combined with a multi-layer x-ray absorption correction model for quantification, and by comparing the x-ray radiation obtained from the quantum well with a reference 10-nm-wide structure, we demonstrate that the quantitative impact of beam propagation on chemical composition can be precisely accounted for, resulting in a composition sensitivity at the atomic scale as low as $\pm 0.25$ at.%. Finally, practical aspects to achieve this high precision level are discussed, particularly in terms of inelastic multislice simulation, uncertainty determination, and sample quality.

# 1. Introduction

Understanding the relationship between atomic-scale structure and macroscopic properties of materials is a crucial step to enable the development of advanced nanotechnologies. To directly probe the chemical properties of solids with ultimate spatial resolution, atomic-scale energy-dispersive x-ray (EDX) spectroscopy performed within a scanning transmission electron microscope (STEM) has become an increasingly popular technique. The first demonstrations of such analyses were reported nearly 15 years ago [1,2], and their efficiency has since advanced significantly. These improvements are attributed to enhancements in microscope instrumentation, particularly the routine use of spherical aberration corrector within a STEM for generating sub-angstrom electron probes and the development of more efficient x-ray detectors, which have significantly increased x-ray detection capabilities.

Since then, the ability to convert the detected x-ray signal into a quantitative measurement of chemical species has been intensively studied. For a long time, the most effective approach for standard nanometer-scale STEM-EDX quantification was based on the relative comparison of different x-ray emissions from the material using the *k*-factors method [3]. This technique directly compares the relative probability of two different atomic transitions to produce x-rays during their de-excitation processes. More recently, the technique has been improved through the incorporation of the so-called $\zeta$ factors [4]. In this approach, each x-ray line is calibrated and treated independently of other radiations. This enables a direct correlation between the x-ray intensity and the local mass-thickness of the material, which is of much importance to account for the absorption effects that affect these radiations by the sample itself. This method has demonstrated remarkably accurate results to measure local chemical composition, achieving relative uncertainties as low as 5% or even less, while preserving the excellent spatial resolution provided by the electron microscope.

Unfortunately, the transition from the nanometer to the atomic scale in STEM spectroscopy introduces several artifacts that strongly limit quantitative analyses. Studying crystalline materials at high resolution in an electron microscope requires observing the sample along a high-symmetry orientation to spatially resolve the atomic columns. In this condition, the projected electrostatic potential of these columns becomes enhanced, significantly impacting the interaction between the electron probe and the material. The increased interaction favors electron propagation along atomic columns and leads to a strong modification of the inelastic signal emission, an effect referred to as electron channeling [5–7]. The modification of both elastic and inelastic interactions under these conditions has been widely discussed, particularly to understand the role of experimental parameters such as sample thickness and probe characteristics on spatial resolution [8–12]. The study of the variation of x-ray emission as a function of the sample's crystallographic orientation also opened the possibility to identify atom positions within the crystal lattice, a technique known as Alchemi [13–15]. The increase in x-ray production under channeling conditions is often a benefit to compensate the low detection yield of this radiation in STEM experiments. However, the major drawback is that each atomic species exhibits different sensitivities to this channeling effect, as the influence on electron wave propagation is strongly dependent on the atomic number [6,16]. Hence, the linear relationship between a material chemical composition and the emitted x-ray intensities under electron irradiation - a fundamental assumption in quantification models – is no longer valid [4,17].

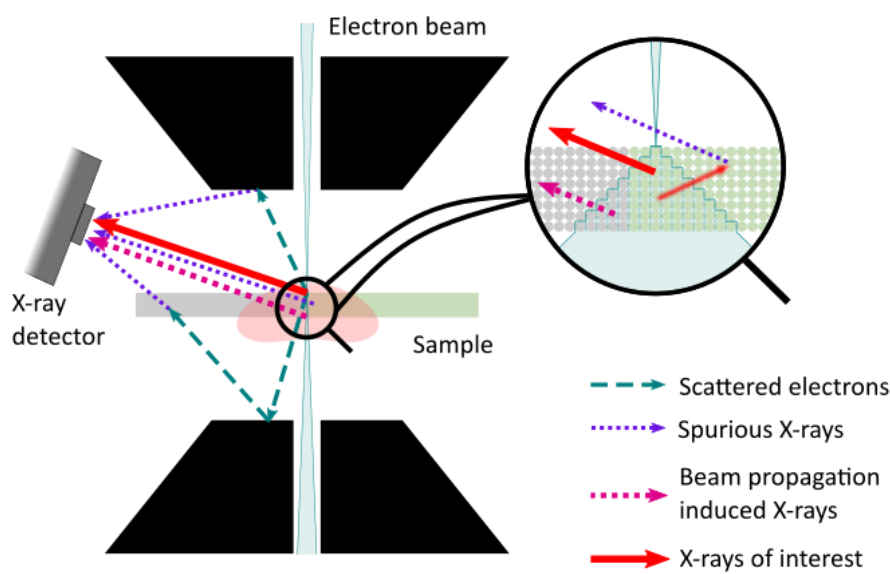

**Fig. 1. Origin of spurious x-rays in the scanning transmission electron microscope (STEM) overlaid with the signal of interest.** The electron beam (light blue) is focused near a crystalline interface within the TEM sample. The x-ray detector collects x-rays emitted from the region of interest (red arrows), but also detects spurious x-rays originating from the chamber and secondary fluorescence (purple arrows) or from the electron beam propagation (magenta arrows).

Another major limitation of high-resolution spectroscopic analysis is related to the increase in the probed material volume due to beam propagation. The probe inevitably diverges as it propagates through the sample since the beam has a certain incident convergence angle, and the surrounding atomic columns act as attractive centers during the scattering processes (referred to as cross-talk) [18,19]. In nanostructured materials, this spatial "sprawl" of the electron beam can excite atoms of a different nature compared to those in the columns located under the probe [20], as illustrated in **Fig. 1**. The schematic illustrates the different sources of detected x-rays, highlighting the potential overlap of x-rays of interest (thick red arrow) with unwanted ones. In this work, we distinguish between two types of undesired x-rays:

(i) spurious x-rays, emitted by scattered electrons far from the zone of interest and ionization processes within the microscope chamber (also reported as stray radiation [21]) or from secondary fluorescence emission (purple arrows);

(ii) beam-propagation induced (BPI) x-rays, coming from the excitation of surrounding atoms within the sample (pink arrow).

Obviously, both signals are highly detrimental to high-accuracy chemical quantification, and must be carefully considered.

To investigate the impact of these effects on quantitative atomic-scale EDX analyses, several studies have highlighted the advantages of performing inelastic multislice simulations. The main advantage of this method lies in its capacity to distinguish signals originating from the actual material structure from those influenced by the aforementioned artifacts, particularly channeling-induced x-ray overproduction and BPI x-rays. Unlike most electron wave propagation simulation algorithms, high-resolution elemental mapping requires the incorporation of inelastic scattering processes. Allen *et al*.

proposed integrating these effects into a dedicated multislice algorithm called μSTEM [22], which incorporates a quantum excitations of phonons (QEP) model [23]. By comparing experimental and simulated data, MacArthur *et al.* demonstrated that channeling effects could be significantly reduced by tilting the sample in a direction where atomic-scale information is not required [12,17]. Simulations have also been successfully used to better investigate the chemical compositions at interfaces, particularly by revealing the effect of sample thickness on the actual origin of the EDX signal [20,24]. Additionally, Chen *et al.* demonstrated the robustness of this approach, showing that the x-ray production yield could be quantitatively predicted if experimental parameters – such as electron probe characteristics, sample orientation, or detector positions – are precisely defined [25,26]. Collectively, these examples, conducted on model systems, demonstrate the efficiency of the simulations in elucidating the effects of beam propagation on experimental results.

Despite significant advancements, a comprehensive understanding of beam-induced artifacts in atomic-scale EDX quantification, particularly within complex device structures, remains incomplete. This study aims to address this gap by experimentally demonstrating the impact of these artifacts on the quantification of practical heterostructures at atomic resolution. We focus on the application of inelastic multislice simulations to accurately identify the origin of collected x-rays, thereby enhancing the accuracy of quantitative results. To this end, we investigate a focused ion beam (FIB)-prepared lamella from a molecular beam-epitaxy (MBE)- grown AlGaN/GaN heterostructure. These materials are of significant interest for the development of solid-state UV emitters, such as LEDs and lasers [27]. In these devices, the control of both interface quality and chemical composition in the quantum wells (QW) and barriers are critical, as these properties directly impact the confined energy levels and the recombination of localized charge carriers.

# 2. Method

## 2.1. Sample studied

The present work involves the observation of two samples. The first one, referred to as Large Quantum Well (LQW), consists of an active region with 20 nm wide $Al_{0.1}Ga_{0.9}N$ barriers and 10 nm tick GaN wells. The second sample, designated Thin Quantum Well (TQW), presents the same structure but with thinner barriers (10 nm) and especially much thinner QWs (1.5 nm). The samples were grown by plasma-assisted molecular beam epitaxy (MBE) on a self-standing GaN substrate, along the c-direction of the wurtzite structure. Readers can refer to previous publications for more details about the growth and the complete structure of the devices [28–30]. The specimens for STEM observations were prepared using a Ga+ focused ion beam (FIB) in a FIB-SEM Zeiss Crossbeam 550 microscope along the [11-20] direction employing the standard *in situ* lift-out technique where lamellas are attached to the posts of lift-out TEM copper grids. To prevent any shadowing effects of the x-ray detectors caused by the grid support, the samples were positioned on the front side of the TEM grid posts after removing the beveled part of their sidewalls. Only a limited area of the lamella was thinned using a progressively decreasing beam voltage, from 30 kV down to 2 kV, to maintain a relatively low sample thickness (<40 nm). The precise thickness was measured by STEM-EELS using the log-ratio method [31], along with an estimation of the inelastic mean free path (λ) based on the Iakoubovskii model [32]. Supplementary Material A outlines the key steps of this preparation procedure and provides a detailed characterization of the sample geometry and thickness maps. The TQW sample analyzed was estimated to be approximately 38 nm thick. In section 4.3, two regions of the LQW sample are analyzed: one with a thickness similar to that of the TQW (~38 nm) for consistency, and another, thicker region, measured at ~133 nm.

## 2.2. EDX analysis

STEM-EDX analyses were performed on a probe-corrected Thermo Fisher Scientific Titan Themis microscope operating at 200 kV, equipped with a Super-X detector system [33]. The system comprises four 30 mm$^2$ windowless silicon drift detectors integrated into the pole piece. The detectors are symmetrically placed at a 45° azimuth angle from each other relatively to the beam axis, with an elevation angle of 18°, enabling a collection efficiency of 5.1 ± 0.5% assuming a solid angle of 0.64 ± 0.06 srad [34]. The beam convergence angle was set around 15 mrad, to maximize spatial resolution while reducing channeling effects (see **section 4.4)**.

The primary challenge in obtaining high-resolution hyperspectral EDX (HR-EDX) data stems from the notable sample drift during the acquisition. To achieve sufficient x-ray production during analysis, the operator has three options: preparing a thicker sample, increasing the beam current (while assuming a constant probe size), or exposing the material to the electron beam for a longer duration. The first option cannot be considered for quantitative HR-EDX analysis, as demonstrated later in **section 4.3**. The second option poses a risk of causing significant damage to the sample, compromising its pristine microstructure. Therefore, the remaining approach is to extend the acquisition time. In a conventional raster-scanning approach, simply increasing the dwell time is not adapted, as the sample may either drift significantly or sustain damage under prolonged beam exposure. For these reasons, we opted for a multi-frame spectrum-imaging (MFSI) approach, consisting of a frame averaging of multiple SI datasets, each acquired in a couple of seconds [35]. This method distributes the total beam dose across multiple frames, substantially lowering the dose rate measured in $e^-/Å^2/s$. SI data were acquired using the Velox® software, and realignment and summation of the SI stack were performed using the simultaneously acquired HAADF signal. This process was automated with a custom Python script based on the scikit-image library [36]. By maintaining a short dwell time (8-16 µs) and limiting the map size to 256 x 256 pixels, individual frames could be recorded in less than a second. This approach minimized frame-to-frame drift, allowing the use of a simple rigid-frame registration procedure. To optimize the count statistics while preventing signal loss due to detector dead time, the probe current was set to 65 pA, and the detector pulse processing time was fixed at 2.9 µs, below the dwell time per pixel. These conditions, combined with the relatively low count rates (on the order of a few kcps), minimize the risk of losing x-ray events [37].

To further enhance the signal-to-noise ratio (SNR) of x-ray intensities, a procedure was applied to periodically project the summed SI data along the growth plane direction. x-ray peak intensities were extracted using the Gaussian fitting method provided by the Hyperspy library [38], modeling the background as a local linear function under each peak. This assumption is only valid due to the reduced sample thickness prepared, which limits absorption of low-energy bremsstrahlung radiation - particularly relevant for the N K transition.

### 2.3. EDX quantification

To convert the detected EDX intensity into a quantitative measurement of the chemical composition, the $\zeta$-factor approach developed by Watanabe and Williams was applied [4]. This method introduces $\zeta_i$ as a factor that relates the detected x-ray intensity $I_i$ to the weight fraction $C_i$ of element $i$ and the specimen's mass-thickness $\rho t$ ($\rho$ and $t$ being the specimen's density and thickness, respectively):

$$C_i = \zeta_i \frac{I_i\, A_i}{\rho t\, D_e} = \zeta_i \frac{I_i e}{I_b \tau} \frac{A_i}{\rho t} \qquad 1$$

Here, $A_i$ represents the absorption correction term for a specific x-ray line $i$. As shown in **Eq. 1**, the beam dose needs $D_e$ to be carefully measured during the acquisition and is determined using the beam current $I_b$ (in pA), the dwell-time $\tau$ (in s) and the electron charge $e$. The $\zeta$ factor depends on the element properties, the x-ray transition and the detectors configuration, but remains independent of experimental parameters such as beam current or exposure time. This characteristic simplifies the analysis and facilitates the comparison of x-ray yields across different samples or thin foils using a single set of $\zeta$ factors, each associated with a specific x-ray transition.

In the case of a parallel-sided lamella, the absorption coefficient $A_i$ is given by [4]:

$$A_i = \frac{(\mu/\rho)_i\, \rho t\, \sin^{-1}\alpha}{1 - \exp[-(\mu/\rho)_i\, \rho t\, \sin^{-1}\alpha]} \qquad 2$$

where $\alpha$ is the x-ray takeoff angle and $(\mu/\rho)_i$ is the mass absorption coefficient for the x-ray radiation generated by element $i$ within the material. Since this coefficient depends on the specimen's mass-thickness, an iterative procedure is employed to achieve convergence on both the concentration and the mass-thickness measurements.

Finally, the uncertainties associated with the measured elemental concentrations ($\Delta C_i$) and sample thickness ($\Delta t$) were calculated following the procedure described by Watanabe and Williams [4]. These uncertainties arise from various sources of error, including the uncertainties in the measured x-ray intensities ($\Delta I_i$) - specifically for Al K, Ga K, and N K - and the uncertainties in the ζ-factors ($\Delta \zeta_i$). By systematically varying these input values and repeating the ζ-factor iterative procedure, the resulting deviations in composition and thickness can be used to estimate the corresponding uncertainties:

$$\Delta C_i = \sqrt{\sum_{k=1}^{n}[C_i(\Delta\zeta_k) - C_i]^2 + \sum_{k=1}^{n}[C_i(\Delta I_k) - C_i]^2} \qquad 3$$

$$\Delta t = \sqrt{\sum_{k=1}^{n}[t(\Delta\zeta_k) - t]^2 + \sum_{k=1}^{n}[t(\Delta I_k) - t]^2} \qquad 4$$

As a result, the set of N data points (or Voronoi cells identified around each atomic column, see **section 3.2**) are associated with non-uniform uncertainties, since these are directly related to the local x-ray intensities. Therefore, when calculating the mean concentration of the i-th element within a specific region of the sample (e.g., in the barriers or the well), we use a weighted mean and the corresponding weighted uncertainty [39,40]:

$$\bar{C}_i = \frac{\sum_N C_i/\Delta C_i^2}{\sum_N 1/\Delta C_i^2} \qquad 5$$

$$\Delta \bar{C}_i = \left(\frac{1}{\sum_N 1/\Delta C_i^2}\right)^{-1/2} \qquad 6$$

# 3. Results

## 3.1. Evidencing the impact of electron beam propagation on EDX intensities

The active region structure of LQW and TQW samples (described in **section 2.1**) were analyzed using high-resolution STEM-HAADF imaging, as shown in **Fig. 2a**. Variation in alloy composition between the wells and barriers produce distinct Z-contrast across the structure, even for relatively small differences in chemical content (only 5 at.% aluminum in AlGaN barriers, which appear noticeably darker than the GaN QWs). **Fig. 2b** compares representative net EDX spectra (counts/s/pA) measured at the center of the GaN wells for both structures, focusing on the Al K transition peak. The spectra were obtained by projecting the maps along the plane growth direction and summing the signal over a 1 nm-wide at the middle of each layer. A significant Al K emission intensity is detected within the GaN well of the TQW structure, whereas the LQW spectrum exhibits a peak barely distinguishable from the background noise. To explain these results, we suggest closely examining the trajectory of the electron beam as it propagates through the two structures. In the case of the LQW sample, the relatively wide

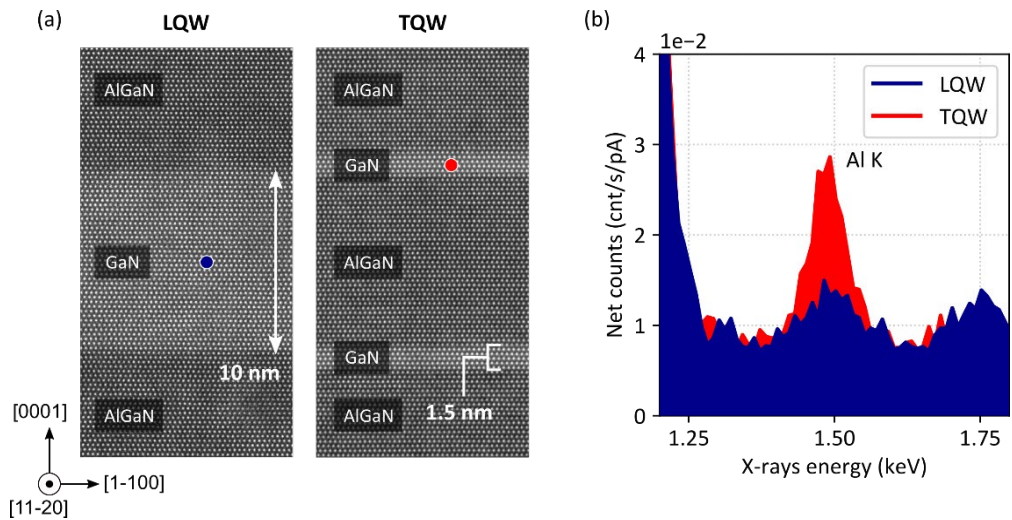

**Fig. 2. Comparison of the EDX Al K emission within the GaN layer for the two QW structures.** (a) STEM-HAADF image of the two samples, indicating the probe location at the QW centers (red and blue dots) during EDX measurements. (b) EDX spectra comparison around the Al K peak between LQW and TQW structures, highlighting the significant increase of aluminum intensity in the thinner GaN layer.

GaN wells (~10 nm) ensure that most electrons interact exclusively with pure GaN material, which is a common assumption in STEM-EDX analyses and simplifies interpretation. However, for the ~ 1.5 nm wide GaN layers in the TQW sample, the situation becomes more complex, as the electron wavefunction intensity may extend beyond the well's interfaces and spreads into the adjacent AlGaN barriers. Consequently, when the beam is focused on a GaN layer, it can still interact with aluminum atoms in the surrounding AlGaN layers, leading to an artificial Al K x-ray signal. This effect presents a significant challenge for compositional analysis, as it becomes difficult to differentiate between genuine aluminum incorporation in the GaN wells and the spurious signal caused by beam propagation effects.

To evaluate the extent of beam broadening in our heterostructures, we performed multislice simulations of electron propagation through the sample thickness along the [11-20] direction using the µSTEM algorithm developed by Allen *et al.* [22]. The simulations were carried out on a super-cell structure containing a single 1.5 nm-wide GaN QW (i.e. comprising 4 atomic planes in the [0001] direction) embedded between two AlGaN barriers, emulating a STEM analysis of the TQW structure. The real-space simulation area was set to a 17 nm square with a maximum grid resolution of 4096 pixels. Particular attention was paid to prevent artifacts resulting from a limited wave vector during computation - further details can be found in **section 4.2**.

**Fig. 3a** shows the evolution of the electron wavefunction intensity $|\varphi|^2$ at three different depths within the TQW crystal, with the beam positioned centrally in the layer, equidistant between two Ga planes. A portion of the super-cell structure used for the simulation is depicted in **Fig. 3b**. The QW boundaries are indicated by white dashed lines in **Fig. 3a**, where it can be observed how the electrons begin to scatter into adjacent layers. Initially, the beam profile resembles the typical Airy function formed by a circular aperture, with nearly 99% of the beam intensity confined within the QW (first panel). As the beam propagates deeper, the electron wave progressively spreads into neighboring atomic columns. To better visualize this effect, **Fig. 3c** quantifies the fraction of the beam contained within the QW as a function of the sample thickness traversed by the electron wave. The black and green curves correspond to the electron beam positioned on a Ga atomic column (labeled as "Ga") and positioned between atomic columns (labeled as "BC"), respectively. For the BC position, almost 90% of the beam remains within the QW even after propagating 40 nm, of the sample, indicating minimal interaction with surrounding layers. In contrast, when the beam is centered on a Ga atomic column (green cross in **Fig. 3b**), the Ga column acts as a strong scattering center, resulting in significantly greater beam broadening; at 40 nm depth, 50% of the electrons have into the AlGaN barriers. These two beam positions account for the minimal and maximal range of scattering effects expected, while other positions (e.g. on a N atomic column) exhibit intermediate broadening behavior. Therefore, this simulation corroborates our previous hypothesis, confirming that beam broadening significantly affects the spectroscopic results in the wells of the TQW structure.

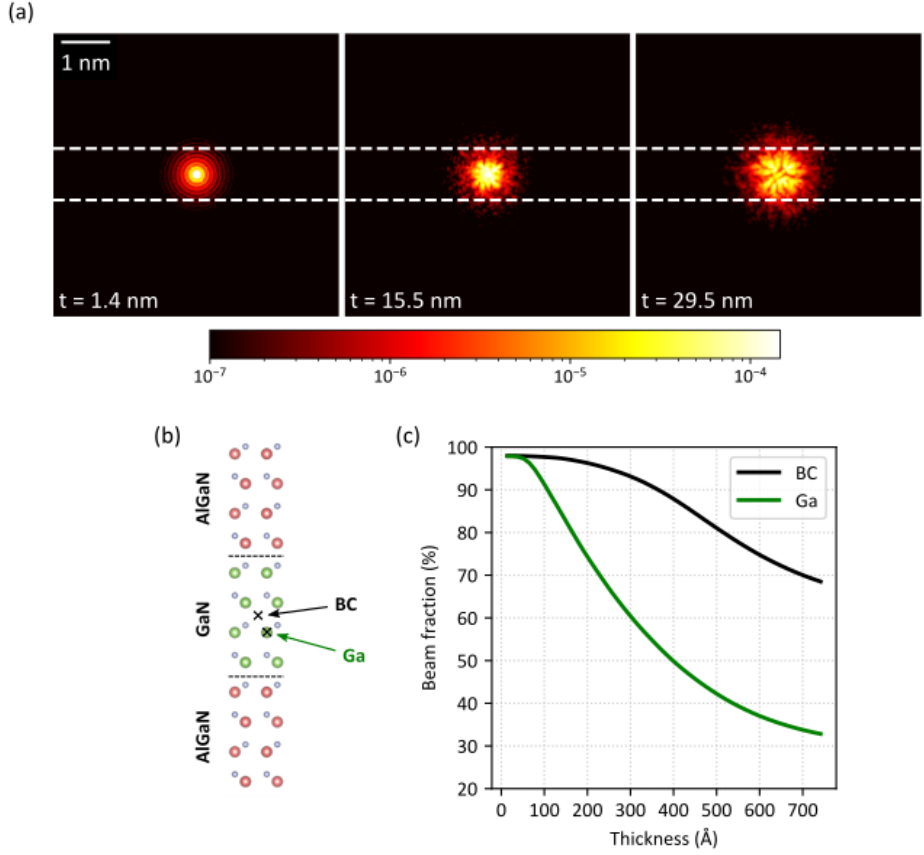

**Fig. 3. Inelastic multislice simulation of probe propagation within a 1.5 nm wide GaN QW in the TQW structure.** (a) Electron wavefunction intensity $|\varphi|^2$ in 3 different crystal depths $t$, with the beam focused at the center of the GaN QW. White dashed lines indicate the GaN QW boundaries. The colors are logarithmically scaled to capture the full beam contributions. (b) Schematic of the TQW simulated structure, showing the two beam positions labeled "Ga" (focused on a Ga atomic columns) and "BC" (focused between atomic columns). (c) Evolution of the beam fraction contained within the GaN QW as a function of crystal depth for "Ga" and "BC" beam positions.

A similar approach was employed in a pure GaN material to ensure that such an effect could not arise in the 10 nm wide QWs of the LQW structure (**Supplementary Material B**). The results confirm that those GaN wells are large enough to prevent electrons from interacting with the surrounding barriers when the beam is positioned in the center of the well.

**Tab. 1** summarizes the beam size extracted from this simulation, depending on the beam position (taken from **Fig. 3b**) and sample thickness (38 and 133 nm). In the case of a thin sample, 60% of the probe is contained within a diameter ranging from 7.9 to 24.4 Å, depending on the beam position. For the Ga position (i.e., with larger scattering amplitude), 90% of the probe is contained within a diameter of less than 7 nm. At this depth (corresponding to the thickness of the samples), it is confirmed that the majority of the probe does not propagate with sufficient amplitude to reach the AlGaN barriers located a few tens of ångströms from the center of the GaN layer in the TQW structure. However, the same conclusion is no longer valid for a sample with a more standard thickness for EDX analysis (in this example, 133 nm), where 60% of the probe is contained within a region of 9 nm in diameter.

These simulations are compared with a commonly used analytical approach as defined by Goldstein *et al.* [41,42], which defines the surface containing 90% of the beam intensity:

$$b_{90} = \frac{Z\,F}{5\,E_0} \sqrt{\frac{\rho}{A}}\, t^{3/2} \qquad 7$$

where $F$ is relativistic correction factor, Z the atomic number, A the atomic mass, and $\rho$ the material density. For compounds such as GaN, we used an effective atomic number $Z_{eff}$ as defined by the Lenz model as $\sum_i f_i Z_i^{1.3} / \sum_i f_i Z_i^{0.3}$ where $f_i$ is the atomic fraction of species $i$. Values for the thin (38 nm) and thick (133 nm) samples are listed in **Tab. 2**. The results are in very good agreement with the multislice simulations and provide a good estimate of beam broadening in GaN crystals. However, the simulations better capture scattering differences between atomic columns, which is important for HR-EDX analyses. The effect of sample thickness on these analyses is discussed further in **section 4.3**.

| Sample thickness (nm) | | 38 | | 133 | |
|---|---|---|---|---|---|
| | Beam fraction | 60% | 90% | 60% | 90% |
| **Multislice simulation (Å)** | Ga column position | 24.4 | 69.4 | 86.4 | X |
| | Between column (BC) position | 7.9 | 16.0 | 22.7 | 105.0 |

**Tab. 1.** Simulated probe diameter (in Å) from multislice calculations as a function of GaN crystal thickness for the two beam positions shown in **Fig. 3b**. Diameters correspond to the regions containing 60% and 90% of the total beam intensity, respectively.

| Sample thickness (nm) | 38 | 133 |
|---|---|---|
| $b_{90}$ (Å) | 22.6 | 148.0 |

**Tab. 2.** Calculated probe diameter $b_{90}$ (in Å) representing the diameter containing 90% of the beam intensity, as defined by Goldstein *et al.* [41,42].

Despite these evidences of a negligible effect of beam propagation inside the well of LQW sample, it is worth noting in **Fig. 2b** that a weak yet noteworthy Al K intensity is measured. This residual radiation is attributed to the production of spurious emission of Al K emission, an effect previously mentioned in the introduction (illustrated in **Fig. 1** by the pink arrows). To the authors' knowledge, the only possible source of aluminum in the microscope (i.e. external to the sample), assuming that microscopists do not use any aluminum-containing ring or clip in the sample holder, comes from the contact used to collect the charge in the EDX detectors. This contact could emit radiation through fluorescence, but the extent of such emission is highly limited. We then deduce that the primary contribution of this spurious signal comes from the sample itself, more specifically through indirect (e.g. from high angle scattered electrons) and/or x-ray secondary fluorescence emission of the aluminum-rich (~5 at%) barriers composing the active region of the structure [29,30]. Although small in the present case, one can expect a much larger effect if these surrounding layers were richer in aluminum. For this study, we take advantage of the possibility to compare the emission between two similar specimens, where the main difference lies in the layer widths. Note that the cladding layers, however, differ as well with more aluminum-rich materials incorporated into the TQW structure, but we can assume that this contribution is minor. In the following, this spurious signal recorded in LQW structure will be used to accurately remove its contribution from the quantitative analysis of the TQW structure in the following parts.

In summary, the results presented in this section demonstrate both experimentally and through simulations the significant impact of electron beam broadening on EDX measurements for nanometer-scale layers. In the following section, we aim to evaluate the consequence of such artifacts on the quantitative analysis of the heterostructures.

### 3.2. Standard HR-EDX quantification of nitride heterostructure

In this section, we determine the Al content along the 1.5 nm GaN QW profile of the TQW structure solely based on the experimental EDX data, as commonly performed in the literature. Therefore, this approach does not account for any beam propagation effects, meaning that we assume that the result obtained at each electron beam position directly corresponds to the composition at its specific location. For that purpose, an HR-EDX mapping was acquired on a single GaN quantum well of the TQW structure. **Fig. 4** shows the results after lattice projection for the three selected atomic transitions (N K, Ga K and Al K). The original intensity maps are provided in **Supplementary Material C**. The clear depletion of Al K signal helps identify the presence of the QW at the center. As expected, the Ga K map exhibits the highest signal-to-background ratio (SBR), primarily due to its greater weight fraction in the alloy and the high yield of x-ray production of this transition. Despite the higher noise level observed for N and Al atoms, our optimized acquisition and data processing procedures are sensitive enough to resolve the atomic column positions for these elements. On the right side of the maps, the application of a Fourier-space filter enhances the contrast between atomic columns and the background, revealing the consistent positions of the respective atom groups. In **Fig. 4**, the Al K signal, previously detected at the center of the QW in **Fig. 2b**, is now spatially resolved through the atomic planes in the HR-EDX mapping. Interestingly, a faint yet distinguishable contrast of the Al-K signal is observed within the QW at the positions of the group III atoms. As we already discussed, this contrast does not necessarily confirm the presence of aluminum within the QW. Indeed, the simulations in **Fig. 3b** suggested that the beam propagation into the barriers is stronger when the probe is focused on a Ga column. As a consequence, an increased production of artificial BPI Al K signal on nominally Al-free Ga columns is expected. Therefore, the experimental results are insufficient to conclusively establish the presence or absence of aluminum within the QW.

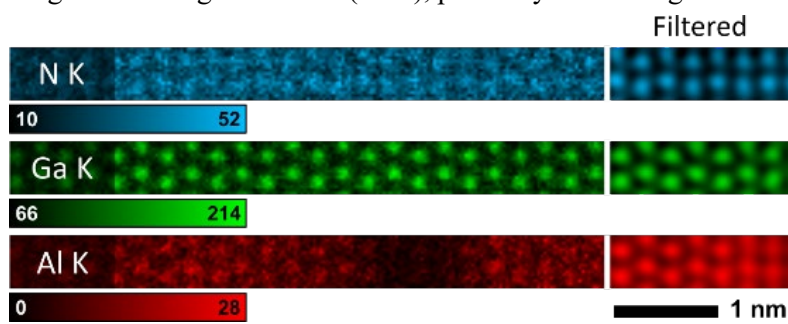

**Fig. 4. Experimental HR-EDX intensity map of a single QW in the TQW structure, displaying the spatial distribution of Al K, Ga K and N K intensities.** The right side of the three maps is Fourier-filtered to emphasize the atomic column positions.

To measure the impact of this signal on composition determination, the $\zeta$ factors approach presented in **section 2.3** was applied. Given the low x-ray signal intensities inherent to atomic-scale EDX mapping, particular care was taken to minimize any potential uncertainties in the quantification process. First, we performed a direct experimental calibration of the $\zeta$ factors inside the microscope using FIB-prepared lamellae of pure GaN and AlN reference samples. Details of this procedure, along with a theoretical description of the $\zeta$-factors, are provided in **Supplementary Material D**. This approach significantly reduces uncertainties related to these factors values, provided the composition and thickness of the reference specimens are accurately known [4]. Since pure AlN and GaN are stochiometric, precise thickness measurements of the lamellae are especially critical to minimizing relative errors. **Tab. 3** presents both experimental and theoretical values obtained for the three $\zeta$ factors of interest. For nitrogen, present in both reference alloys, the experimental calibration shows excellent reproducibility, with similar factor values obtained from GaN and AlN measurements. The significant discrepancies observed between theoretical and experimental $\zeta$ factors (except for N K) highlight the importance of experimental calibration for improved accuracy.

Another critical parameter in the quantitative procedure is the correction of x-ray absorption. The approach proposed by Watanabe and Williams [4] is highly effective for studying samples that are uniform and homogenous over several hundreds of nanometers - corresponding to the lateral trajectory of x-rays before exiting the sample and reaching the detectors. However, for heterostructures, the material density $\rho_i$ encountered by x-rays during propagation may differ from the material density $\rho$ where the x-rays are initially generated. This aspect - particularly significant in the case of EDX tomography [43,44] – becomes especially relevant when the region of interest is located near the sample surface. In such cases, the presence of vacuum or FIB protective layers drastically alters the radiation absorption effect. Given the prior knowledge of the structure and composition of the layers surrounding the region of interest, the iterative procedure of the $\zeta$ factors approach was adapted to incorporate a multi-layer absorption correction model.

| Transition lines | | Al K | Ga K | N K |
|---|---|---|---|---|
| **Experimental $\zeta$ factors (kg/m²)** | AlN | 109.2 ± 4.4 | / | 119.2 ± 3.1 |
| | GaN | / | 216.6 ± 5.6 | 121.7 ± 3.7 |
| **Theoretical $\zeta$ factors (kg/m²)** | | 104.4 | 226.2 | 120.2 |

**Tab. 3. Experimental and theoretical $\zeta$ factors for the Al K, Ga K and N K transitions.**

The above quantitative procedure is applied to the projected profile of the data cube along the growth plane direction. To ensure sufficient SNR for the measurement of chemical composition, we consider the sum of x-ray intensity for each atomic plane along the c-axis growth direction, in a similar manner to the usual Voronoï cell approach used in quantitative HAADF [45]. Note that, as the N and III-metals atomic planes are slightly shifted along the growth direction, they were considered in the same atomic plane to obtain the final quantification result. A more detailed discussion on the number of x-ray counts required for quantification, justifying the Voronoï approach, is given in **section 4.1**. The resulting chemical concentration profile of Al, Ga and N across the QW is presented in **Fig. 5**, giving the composition for each atomic plane in the c-direction. Excellent results are obtained, with a mean nitrogen content of 49.2 ± 0.6 at.% along the whole stack profile, close to the stochiometric composition of these alloys. Approximately 5.3 ± 0.2 at.% of Al in the barriers is measured, in good agreement with the nominal values. The quantified aluminum profile exhibits a progressive transition between the barriers and the well over 2 atomic planes, with a slight difference between the AlGaN/GaN and GaN/AlGaN interfaces. Interestingly, even after accounting for the spurious signal calibrated on the LQW sample (see **section 3.1**), the Al chemical content within the GaN QW does not drop to zero. Considering the QW nominal structure with 4 GaN atomic planes, a mean concentration of 1.0 ± 0.2 at.% is found. This result represents a surprising but significant deviation from the expected value (i.e., Al content of 0 at.%), reinforcing the probable effect of electron beam propagation on the composition analysis of nanometer-scale layers in heterostructures. Without further investigations, any potential variation in chemical composition that could impact the device band properties would remain undetectable.

**Fig. 5** also presents the thickness profile obtained using the $\zeta$-factor method, under the assumption that the material density can be reliably inferred from the simultaneously calculated composition. Throughout the profile, the sample thickness remains nearly constant despite changes in layer composition. The mean thickness is approximately 81 nm, with an uncertainty of about ±6 nm per atomic column (at a 99% confidence level). A slight increase in thickness (~84 nm) is observed on the left side, attributed to a local reduction in nitrogen content likely caused by increased beam irradiation during scan parking. However, the thickness is clearly overestimated compared to the EELS-based measurement, which yielded a value of ~38 nm for the same region. This discrepancy is expected, as x-ray intensities are always significantly enhanced when the crystal is oriented along a high-index zone axis, as is the case during high-resolution imaging. According to the $\zeta$-factor relation (**Eq. 1**), the estimated thickness scales linearly with x-ray intensity. In the present case, the overestimation factor is ~2.1, in good agreement with previous reports of x-ray overproduction under strong dynamical diffraction conditions [46]. The main consequence of this thickness overestimation is an excessive correction for x-ray absorption, particularly for the low-energy N K line. To ensure that this effect does not significantly impact the quantification of nitride materials, we carefully evaluated it and found that, provided the electron beam convergence angle is sufficiently high, the resulting error in absorption estimation leads to a maximum concentration deviation of about 2 at.% for a stoichiometric GaN crystal. A more detailed discussion of this effect is provided in **section 4.4**.

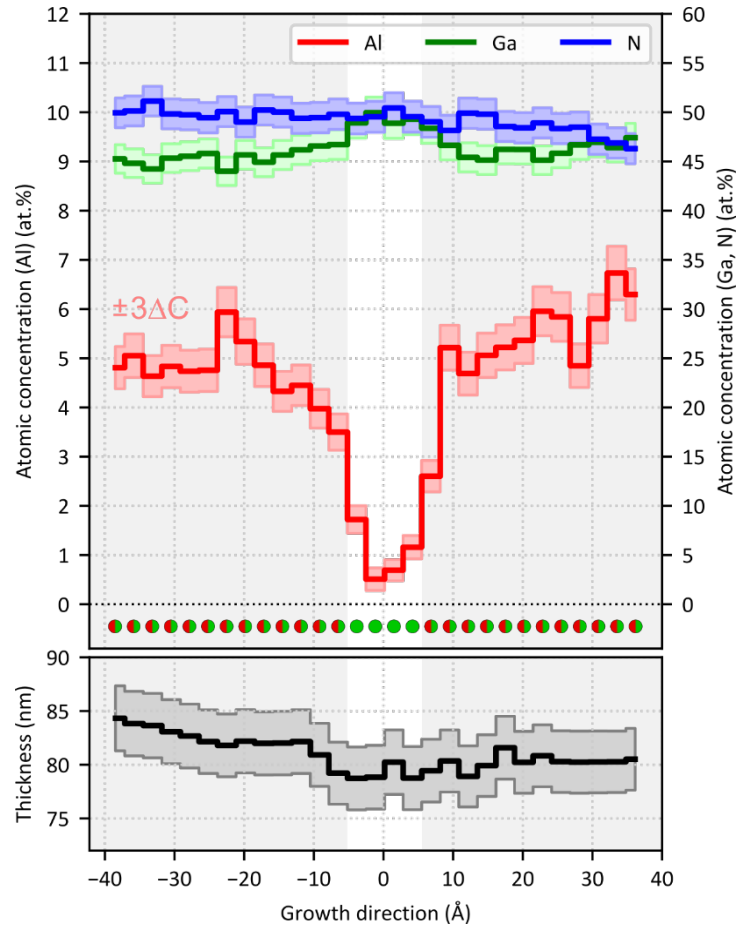

**Fig. 5. HR-EDX chemical concentration profile of Ga, N, and Al across a single quantum well in the TQW heterostructure,** obtained using the standard ζ-factor method without accounting for electron beam propagation. Each step in the graph corresponds to a single atomic plane along the growth direction. The nominal AlGaN/GaN/AlGaN layer structure is shown below the graph and indicated by the background color shading. The lower part of the figure shows the thickness profile of the heterostructure extracted from the ζ-factor analysis.

## 3.3. Incorporating beam propagation effects into HR-EDX quantification

In the following, the impact of beam propagation on composition analysis is considered by conducting simulations with the μSTEM algorithm, which incorporates inelastic interaction and allows for the simulation of the EDX signal of interest [22]. In this optic, the simulations results are compared to the experimental HR-EDX data. **Fig. 6** provides an example of an HR-EDX simulation based on a structural model of the TQW stack, with sharp chemical transitions between the layers and no aluminum incorporated in the well (corresponding to the Al profile depicted in black in **Fig. 7a**). While the simulated map closely resembles the experimentally obtained results (**Fig. 4**), notable differences persist, particularly the absence of any Al K signal at the QW edges or within the central region of the well. Moreover, the contrast between atomic column and background intensity differs between experimental and simulated data. This discrepancy stems from the partial coherence of the electron probe, a widely discussed effect in the literature to explain the difficulty to quantitatively simulate the contrast both in imaging [8,47–49] and spectroscopic [50–52] analyses. To address this, the common approach consists in convoluting the resulting atomic-scale mapping with a certain electron source distribution function [53]. In the present case, the best match between experimental data and simulated results was achieved using a Gaussian function with a full width at half maximum (FWHM) of 0.8 Å, consistent with typical values reported in the literature [8,54].

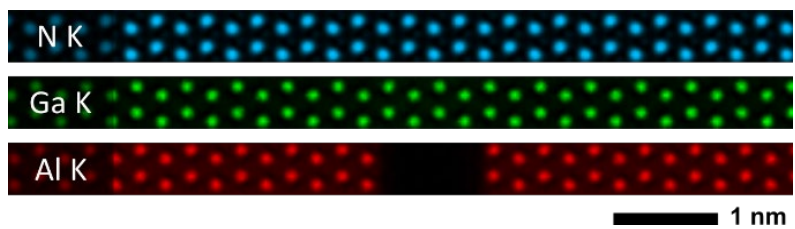

**Fig. 6. Simulated HR-EDX transition intensity maps of an $Al_{0.1}Ga_{0.9}N/GaN/Al_{0.1}Ga_{0.9}N$ QW stack, based on sharp interface model with no aluminum incorporated in the well.**

**Fig. 7b** illustrates the comparison between the experimental and simulated profiles of Al-K x-ray intensity, obtained by projecting the maps along the [1-100] direction. To facilitate a quantitative comparison of x-ray emission yield in the QW, all intensity profiles are normalized with respect to the barriers. This involves comparing simulated and experimental results in terms of the fraction $\delta I_{Al}$ of emitted radiation relative to a zone of reference, i.e. the AlGaN layers of known composition. This fraction is then defined by the relation $\delta I_{Al} = I_{Al} / \overline{I_{Al}^{bar}}$, where $\overline{I_{Al}^{bar}}$ corresponds to the mean intensity measured in the barriers. Unlike comparisons made on an absolute scale [25,26], which require meticulous calibration of microscope settings (e.g. beam current, detector solid angle and efficiency), this relative approach is less sensitive to calibration uncertainties and can be implemented more straightforwardly. However, it requires a reference portion of the sample near the region of interest. In this study, the aluminum content in the AlGaN barriers was consistently determined using both EDX (present work) and x-ray diffraction [55]. Note that the composition measured with conventional EDX is slightly higher than the nominal value (**Fig. 5**). Therefore, we verified that small variations in the reference Al barriers concentration up to $\pm 0.5$ at.% do not alter the conclusion from this work, as shown in **Supplementary Material E**.

As presented in **Fig. 7b** and **c**, the experimental mean value of the Al K intensity within the QW (relative to the mean signal in the barriers) reaches $\delta I_{Al} = 19.1 \pm 1.4\%$. The black curve in **Fig. 7b** corresponds to the simulated Al profile for the previously described sharp interface model without any Al within the well (depicted in black in **Fig. 7a**). The comparison highlights a clear discrepancy between the experiment and simulation results within the QW, particularly in the mean Al K signal intensity. The predicted value from the simulated sharp model does not exceed $\delta I_{Al} = 11.5\%$. The profile presents an almost constant value throughout the QW atomic planes, unlike the pronounced experimental contrast (**Fig. 4**). Furthermore, the simulation does note reproduce the gradual transition experimentally observed between the AlGaN and GaN layers. Consequently, the sharp interface model with no Al within the QW fails to accurately reproduce the experimental data.

To achieve a more realistic Al profile in the structure, we propose modifying two structural properties of the stack: (i) the degree of chemical transition (i.e. in terms of abruptness) at the interfaces and (ii) the potential incorporation of aluminum within the well. The Al gradient is modeled using the second Fick law of diffusion, which describes the concentration evolution between two layers of constant chemical composition. In particular, this model has been used to study annealing effect on nitride heterostructure interfaces [56,57]. The aluminum concentration $C$ of an atomic plane $x_i$ can be expressed as follow [58]:

$$C(x_i) = C_{bar} - \frac{C_{bar} - C_{QW}}{2} \left( \text{erf}\left[\frac{x_i - x_1}{2\sqrt{Dt}}\right] - \text{erf}\left[\frac{x_i - x_2}{2\sqrt{Dt}}\right] \right) \qquad 8$$

where $C_{bar}$ and $C_{QW}$ are the aluminum concentrations in the barriers and QW, respectively, while $x_1$ and $x_2$ denote the interface positions relative to the QW center. The term $\sqrt{Dt}$ corresponds to the diffusion length. It is important to note that, in this study, the diffusion model is not employed to measure any chemical species' diffusion, but rather to describe the gradual incorporation of group III atoms during growth. The diffusion length is then treated as a free parameter to optimize the simulation results, and effectively capture the progressive incorporation of group III atoms during the different growth steps.

A first tested model, only incorporating a chemical gradient, is represented by the blue curve in **Fig. 7a**. An optimal diffusion length of 1.9 Å, smaller than one atomic layer along the c-axis direction, was determined. As shown in **Fig. 7b,** introducing a chemical gradient significantly improves agreement with the experimental data. The evolution of the interface gradient is now accurately captured, with the theoretical curves falling within the uncertainty range of the experimental results. The simulated Al K intensity within the QW rises to $\delta I_{Al} = 15.7\%$, considerably closer to the experimentally established value (i.e. 19.1 ± 1.4%). This intensity, theoretically predicted with no aluminum in the QW, originates from interactions with aluminum atoms within the barriers, and further confirms the significant effect of beam propagation on the EDX results. However, the periodic variation relative to the atomic planes resolved inside the QW is only partially predicted by the simulation. The significant difference in amplitudes suggests the likely presence of genuine aluminum within the well. To test his hypothesis, two additional structural models were considered, fixing $C_{QW}$ (**Eq. 8**) at 0.5 (grey curve) and 1.0 at.% (green curve) of aluminum, respectively. The corresponding chemical profiles are depicted in **Fig. 7a**. As shown in **Fig. 7b**, both models result in a significant increase in the simulated Al K intensity within the QW, while the Al interfacial intensities remain almost unchanged. The simulated relative intensities within the well reach $\delta I_{Al} = 24.0\%$ and 31.8%, respectively. The differences in simulated Al K profiles in the QW are remarkably high, given the relatively small changes in the tested well chemical compositions. While the 1.0 at.% model aligns with the aluminum content measured using the ζ-factors method (**section 3.2**), the predicted intensity is highly overestimating the experimental data, even considering the large uncertainties in the experimental profile. Both models better replicate the periodic variations of the crystalline planes inside the well, although the experimental data still exhibit a surprisingly higher contrast. A fifth model (not shown on **Fig. 7**), incorporating only 0.25 at.% of Al within the QW, yields a simulated ratio $\delta I_{Al}$ of 19.9%, offering the closest match to the experimental results. Within the uncertainties, this simulation-based approach, which accounts for the effects of beam propagation, suggests an aluminum content of 0.25 at.% in this nominally pure GaN layer, with an upper limit of 0.5 at.%. This result significantly differs from the 1.0 at.% of Al measured from the standard EDX quantification approach (**section 3.2**), with the discrepancy attributed to the production of significant BPI x-ray signal.

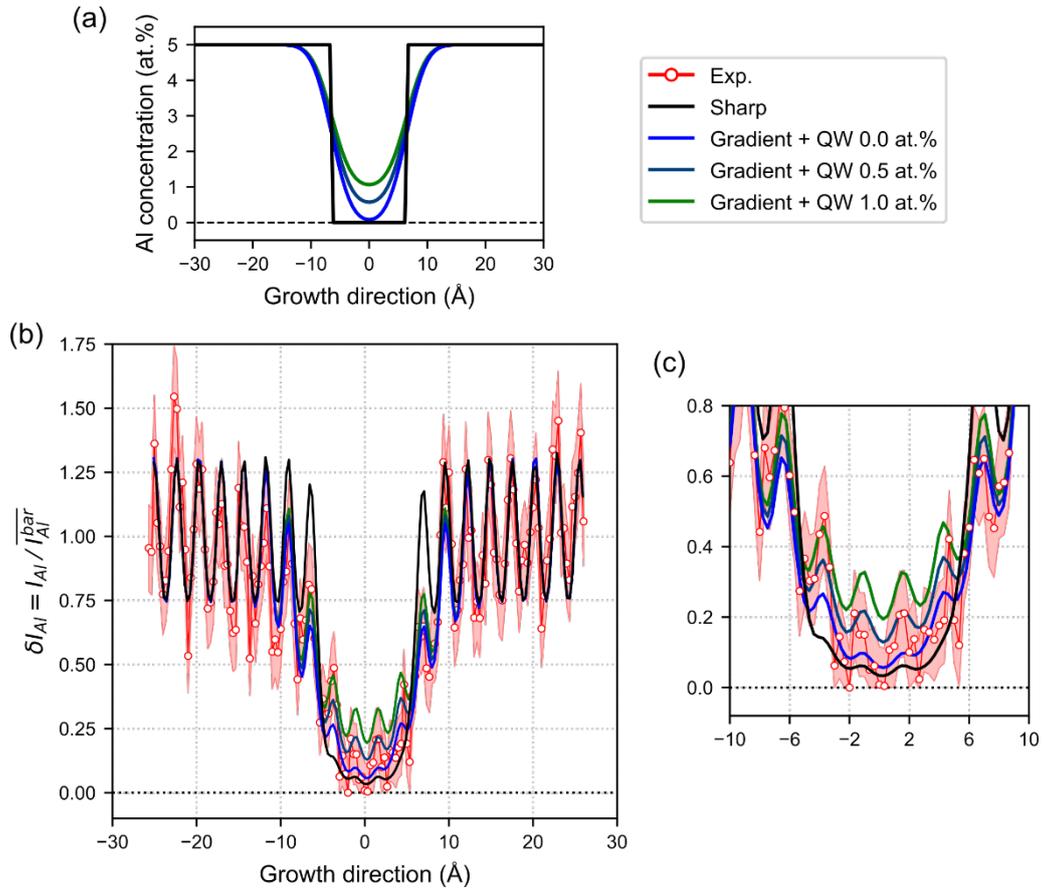

**Fig. 7. Comparison of experimental and simulated Al K intensity profiles based on 4 different QW structural models of the stack.** (a) Al chemical profiles for the different models tested in the simulation. (b) Experimental Al K intensity profile compared to the results of the four simulations based on the different stack structures shown in (a). (c) Focused view on the QW section.

# 4. Discussion

## 4.1. Uncertainty determination in HR-EDX quantification

Since the advent of x-ray spectroscopy generated from the interaction with an accelerated electron beam, one of the primary challenges have been obtaining sufficient signal for quantitative measurements to be meaningful. This problem is particularly significant in STEM, where the reduced interaction volume leads to a lower yield of detected x-rays. Naturally, this issue is exacerbated in high-resolution spectroscopy. In this context, it is essential to establish stringent statistical criteria to objectively assess the reliability of the results obtained.

The key idea is to define the minimum net signal required (i.e., the peak observed above the background) for a chemical element to be considered present. An electronic transition peak is considered discernible (with 99% accuracy) from the background if its net intensity $I_i$ exceeds three times the background noise, i.e., $I_i > 3\sigma_i^B$ [59]. Since x-rays detection follows Poisson statistics, $\sigma_i^B$ is given by the square root of the background intensity $B_i$. The background intensity is determined by extrapolating the values before and after the peak of interest, locally modeled with a straight line, and integrating the intensity over an energy window equivalent to 3 times the FWHM of the peak. Note that for cases of extremely low intensities beyond *bremsstrahlung*, some publications define $\sigma_i^B$ as $\sqrt{2B_i}$ [4,60], which we adopted in the study for consistency with our description of the uncertainty associated with net intensity.

**Fig. 8a** compares the ratios $\Delta_{Al} = I_{Al} / \sqrt{2B_{Al}}$ obtained from the raw intensity profile of the Al K transition (i.e., directly extracted from the projected map in **Fig. 4**, also shown in **Fig. 7b**) and from the application of Voronoi cells employed for the quantification in **Fig. 5**. These cells are generated by spatially integrating the signal of each electronic transition of interest within a limit defined by the interplanar atomic distance along the growth direction. On average, 8 scan points are included in each cell, significantly increasing the number of x-rays while retaining proper atomic-scale information. In the raw profile (purple circles in **Fig. 8a**), the ratio $\Delta_{Al}$ remains above the detectability threshold of 3 (horizontal black line) inside the AlGaN barriers, but drops well below this limit within the QW region. The subtraction of spurious intensity identified in the LQW structure (**section 3.1**) results in very low Al K x-ray intensity in this nominally aluminum-free zone. However, as shown by the continuous stepped curve in **Fig. 8a**, the application of Voronoi cells greatly compensates this statistical limitation. Consequently, all intensities related with each atomic plane exceed the detectability criterion, including the 4 atomic planes belonging to the quantum well.

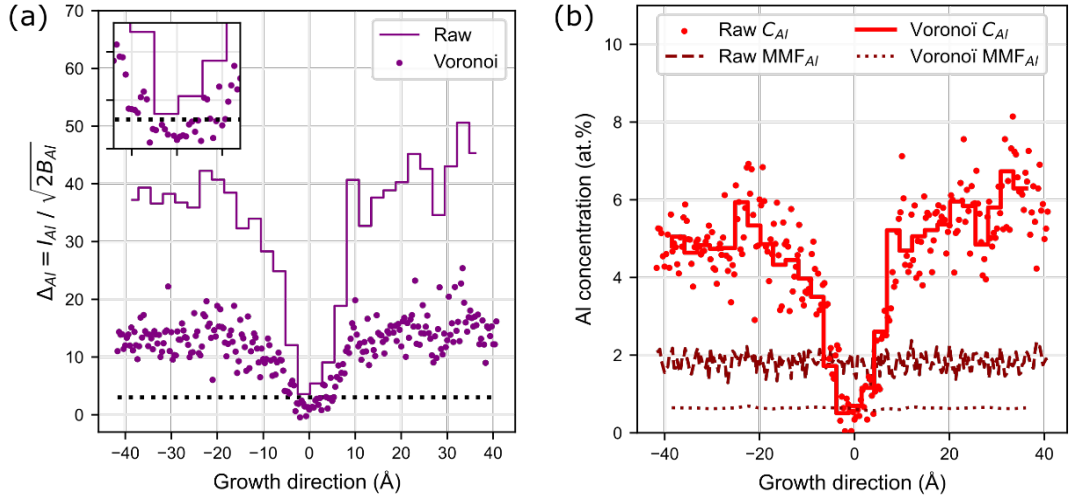

**Fig. 8. Inspection of EDX intensity profile quantifiability, comparing raw data with the application of Voronoi cells.** (a) Evolution of the detectability criterion ($\Delta_{Al}$) relative to the threshold of 3, indicated by the horizontal black dashed line. The inset provides a clearer view of the result within the QW region. (b) Comparison of aluminum quantification profiles ($C_{Al}$) with the minimum mass fraction (MMF) defined in the ζ-factor theory.

To directly connect this detectability criterion to the quantification process, Watanabe and Williams proposed to incorporate this threshold into the theory of ζ factors, to define the minimum mass fraction (MMF) detectable. The MMF represents the lowest concentration of a given chemical element that can be detected in a material [4]. Considering various experimental parameters relevant to the quantification process (probe current, sample thickness, ζ-factor of the element in question) along with the detected continuous background intensity $B_i$, the MMF is expressed as:

$$MMF_i = \frac{\zeta_i 3\sqrt{2B_i}}{D_e \rho t} \qquad 9$$

**Fig. 8b** compares the aluminum quantification profiles (red dots and stepped curve) obtained from the raw intensity profile and after Voronoi cell processing, along with their respective MMFs (brown dashed and dotted lines). The average aluminum MMF value for the raw profile is 1.8 at.%, while the application of Voronoi cells reduces this threshold to approximately 0.6 at.%. Note that the measured Al concentration within the QW (~1.0 at.%) approach this limit, which is consistent with the intensity detectability criterion. This result validates the application of Voronoi cells for the quantification process presented in **Fig. 5**, providing a robust statistical framework that ensures the reliability and significance of the results.

### 4.2. Artifacts in multislice simulation

The approach employed in this study requires careful attention to the determination of simulation parameters, becoming crucial when quantitatvely comparing experimental and simulated data [61]. While no uncertainty in terms of precision is expected from inelastic simulations, this section adresses potential inaccuracies (i.e. deviations from "true" value) arising from numerical artifacts.

In multislice theory, wave propagation calculations rely on the descratization in $N$ pixels of the real-space superlattice of physical size $L$ nm, support of the simulation. During these calculations, Fast Fourier Transform (FFT) operations impose a periodicity condition on this real-space grid. Consequently, any proportion of the beam propagating toward the grid edges is artificially wrapped arround to the opposite edge, potentially causing unintended superposition of the wave with itself. This so-called wrap-around effect [61] is particularly problematic when studying non-hmonenous materials like heterostructures, as it may generate artificial interfaces at the grid edges. To minimize the wrap-around effect, the physical grid size $L$ must be maximized. However, increasing $L$ inversely affects the maximum electron wave vector $\beta_{max}$ that can be considered during computation. This relationship is defined by the limited number pixels in the real-space grid:

$$\beta_{max} = \frac{\lambda}{3}\frac{N}{L} \qquad 10$$

where $\lambda$ is the electron wavelength. To counteract the decrease in $\beta_{max}$ caused by increasing $L$ the number of pixels in the simulation can be increased. However, this adjustment significantly raises computational resource requirements and simulation time (the simulation being conducting in 2 dimensions, this scales quadratically). In this study, the wrap-around was evaluated using different numerical parameters to establish optimal conditions while maintaining reasonable computation times. Two types of simulations were conducted:

- **Point-scan simulation** for the observation of beam propagation, where the electron beam is fixed at a single position (**Fig. 3**).
- **Spatially resolved simulation** to map the x-rays intensity distribution across the heterostructure, with several thousand electron beam positions operated (**Fig. 6**).

The latter simulation type is significantly more computentionally demending and requires a reduction in grid resolution. **Tab. 4** summarizes the parameters used for both simulation types.

| Simulation type | Total grid length $L$ (Å) | Grid resolution (pixels) | Spatial resolution $\Delta x$ (Å) | Maximum wave vector $\beta_{max}$ (mrad) |
|---|---|---|---|---|
| Point-scan (beam propagation) | 176.752 | 4096 | 4.315×10$^{-2}$ | 193.72 |
| Mapping (HR-EDX) | 103.4 | 2048 | 5.05×10$^{-2}$ | 165.5 |

**Tab. 4.** Parameters used for the two types of simulations: point-scan mode (study of beam propagation) and spatially resolved x-ray production (HR-EDX).

Note the reduction in the physical grid length $L$ for the HR-EDX mapping mode to maintain a relatively high $\beta_{max}$ (**Eq. 10**). Under these conditions, less than 5% of the total electron intensity was lost due to the limited wave vector range, confirming the relevance of the chosen simulation parameters.

### 4.3. Thin sample condition

As highlighted in the literature, sample thickness is a key parameter for successfully extracting quantitative information from inelastic simulation [20,24]. Theoretically, experimental data and simulations should yield consistent results if the correct structural model of the sample is used as input. However, uncertainties in the data can limit the precision of structural and chemical information extracted from the sample.

The effect of sample thickness was evaluated by studying the same interface between the well and the barrier in the LQW structure. Two HR-EDX profiles were acquired over a 6 nm region around the AlGaN/GaN interface, using FIB-prepared lamellae with different thickness windows of 38 nm and 133 nm, respectively, as determined by STEM-EELS. The results are presented in **Fig. 9a-b** (thin sample) and **c-d** (thick sample). In a same manner than **section 3.3**, these 2 experimental profiles were compared to simulations two structural models: (i) a sharp interface model and (ii) a gradient interface model incorporating a gradual aluminum transition, as defined by **Eq. 8**. The objective is to evaluate whether the experimental data could distinguish between the predictions of the two models, depending on the sample thickness.

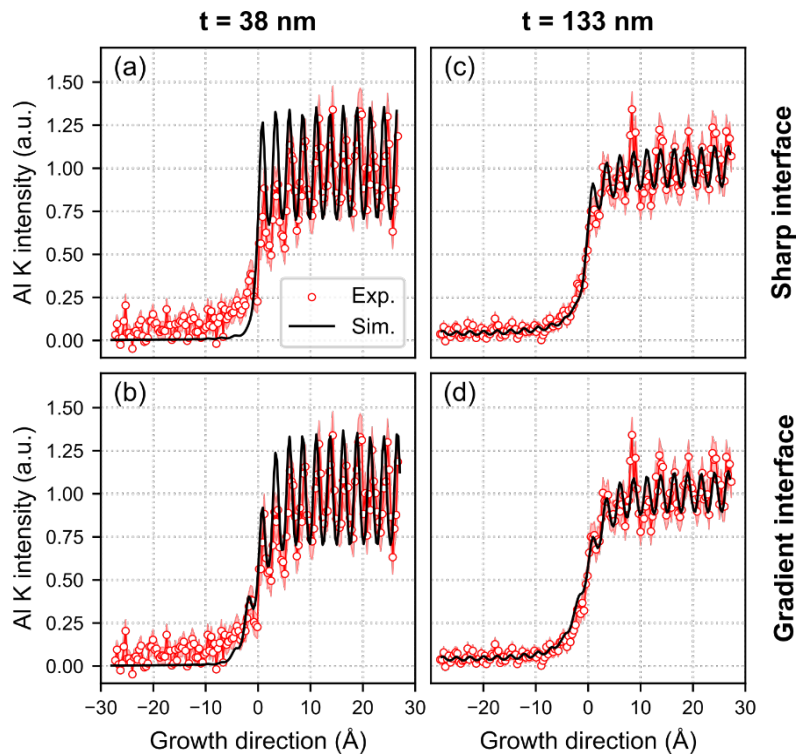

**Fig. 9**. **Impact of sample thickness (38 and 133 nm) on the accuracy of comparing experimental and simulated HR-EDX datasets,** using (a-b) a sharp interface model and (c-d) a gradient interface model.

For both sample thicknesses, the Al K intensity profiles exhibit a clear, gradual transition between GaN QW and the AlGaN barrier (**Fig. 9**), consistent with observations made at the same type of interface in the TQW structure's active region. Within the Al-rich layer, strong intensity variations corresponding to atomic planes are observed in both datasets. However, the atomic column contrast is relatively stronger in the thinner sample [11]. This difference in contrast is remarkably well reproduced by simulations.

The comparison of the two structural models provides rich insights. For the thin sample, the sharp interface model fails to predict the experimentally observed gradual transition of Al K intensity, which spans approximately ~2 nm (**Fig. 9a**). In contrast, the agreement improves significantly when the data are acquired from the thicker 133 nm sample (**Fig. 9b**). Based on this latter comparison alone, one might incorrectly conclude that the sharp interface model accurately describes the experimental results, and that the observed gradual transition is only a consequence of beam propagation effects.

To address this discrepancy, the experimental profiles were compared to simulations incorporating a chemical gradient at the interface. For the thin sample, this gradient-based model achieved a much closer match to the experimental data, particularly for the two atomic planes located just beyond the interface (**Fig. 9c**). However, in the thicker 133 nm sample, nearly no distinction could be made between the predictions of the sharp and gradual interfaces (**Fig. 9d**). Under such conditions, the crystal thickness that the electron wave must travers becomes too significant, causing any variations in structural properties at short scale to be blurred by the beam propagation, even when simulations are applied.

### 4.4. Channeling effects on EDX intensities

As previously mentioned, channeling effects significantly influences x-ray production resulting in higher radiation intensities when the electron beam is aligned with high-symmetry zone axes. This variation depends on several factors, including the specific atomic transitions involved, the material's chemical composition and experimental parameters such as convergence angle [62] and sample orientation [13].

In this study, an initial investigation aimed to identify the experimental conditions under which channeling-induced x-ray overproduction could be minimized and disregarded as a factor affecting the chemical composition measurements. The chemical composition of pure stochiometric GaN layer was determined using the ζ-factor approach while varying the crystal orientation and convergence angle, in order to observe their respective impacts. The results showed that channeling effects can overestimate the gallium concentration by up to 2 at.% when the convergence angle exceeds ~10 mrad (**Supplementary Information F**). The Ga K transitions having a higher energy compared to N K, the smaller impact factor associated will then promote a more important production of x-rays under channeling conditions compared to nitrogen [16]. In AlGaN layers analyzed at high resolution (i.e. along a zone axis) (**Fig. 5**), this effect can account for the slight deviation observed between the measured nitrogen concentration of 49.2 at.% and the expected stochiometric value of 50 at.%. However, when measuring trace aluminum concentrations within the GaN QW, the contribution of this effect is negligible compared to the departure caused by the propagation of the beam in the surrounding layers.

As Watanabe and Williams suggested in their pioneering article [4], incorporating channeling effects into the quantification procedure could offer significant value. This could be achieved either through experimental calibration or simulations to anticipate these effects. However, such an approach would require precise knowledge of the local crystal orientation relative to the zone axis, by combining for instance EDX mapping and Positioned Averaged Convergent Electron Diffraction (PACBED) [63,64].

# 5. Conclusion

In conclusion, this study presents a comprehensive procedure based on the comparison between experimental HR-EDX data and simulations in order to extract structural and chemical information from heterostructures at the atomic scale. The demonstration was done on a GaN quantum well (~1.5 nm wide), whose precise chemical composition and interfacial properties were initially unknown. By testing several structural models of the stack, we successfully separated the contributions of beam propagation and the actual x-ray emission from the material of interest. Our results revealed a certain gradient at the interfaces of the structure, and showed that the small aluminum content (1.0 ± 0.2 at.%) detected inside the well - as measured by standard ($\zeta$-factors) quantification approach - was overestimated by at least a factor of two because of the beam propagation effect. The capability to measure chemical composition with high precision (~0.25 at.%) was demonstrated, which is of great interest to improves the analysis performance of STEM spectroscopy techniques, and more generally to better understand the properties of semiconductor devices mode from these nano-structures.

The limits of this simulation-based approach were explored by focusing on extremely small variations in EDX intensities at the highest spatial resolutions achievable in an electron microscope. Although the observed effects may appear minor, their importance is expected to be more pronounced in crystals presenting closer atomic planes, when using lower electron beam energy or within materials presenting higher Z numbers. With the advent of new-generation EDX detectors, enlarging the amount of collected x-rays and improving the sensitivity of this spectroscopy, the need to account for these effects will become even more critical [65,66].

A key advantage of the simulation-based method is its direct use of x-ray intensity variation through the structure, eliminating the need for a quantification procedure that relies on comparisons between intensities from different atomic transitions. However, one limitation of this approach is the necessity for a precise reference zone with known composition, such as the AlGaN barriers used in this study. To overcome this limitation, comparisons between experimental and simulated data on an absolute scale - i.e. by directly working on the number of x-rays generated – could offer a more robust solution, as demonstrated in previous studies [25,26]. This method provides rich insights, allowing for the quantitative evaluation of experimental parameters, including those related to the microscope and detectors, whose values are often provided by manufacturers. However, this approach requires an accurate knowledge of detection geometry to incorporate these parameters into simulations.

It should be finally noted that only a limited number of structural models were tested as simulation input. Ideally, the heterostructure properties could be refined through on iterative procedure to minimize the gap between experimental and predicted Al K profiles, progressively modifying each atomic column composition until convergence. In practice, due to the significant simulation time required – ranging from a few days to a week depending on sample thickness and chemical homogeneity though the stack - this approach was limited to a manual search for an optimal QW structure, as shown in **Fig. 7**. We expect that better numerical approaches will offer improved performance and accuracy in refining this procedure.

**Declaration of Competing Interest**

The authors have no conflicts to disclose.

**Declaration of generative AI and AI-assisted technologies in the writing process**

During the preparation of this work, the authors used ChatGPT in order to improve the readability and language of the manuscript. After using this tool, the authors reviewed and edited the content as needed and take full responsibility for the content of the published article

**Data availability**

Data will be made available on request.

**Acknowledgments**

This work, carried out on the Platform for Nanocharacterisation (PFNC), was supported by the "Recherche Technologique de Base" and "France 2030 - ANR-22-PEEL-0014" programs of the French National Research Agency (ANR).

# SUPPLEMENTARY MATERIAL

# Impact of electron beam propagation on high-resolution quantitative chemical analysis of 1-nm-wide GaN/AlGaN quantum wells

**Florian Castioni**[1,2], **Patrick Quéméré**[1], **Sergi Cuesta**[3], **Vincent Delaye**[1], **Pascale Bayle-Guillemaud**[3], **Eva Monroy**[3], **Eric Robin**[3], **Nicolas Bernier**[1,*]

[1] Univ. Grenoble Alpes, CEA, LETI, 38000 Grenoble, France

[2] Univ. Paris-Saclay, CNRS, Laboratoire de Physique des Solides, 91405, Orsay, France

[3] Univ. Grenoble Alpes, CEA, IRIG, 38000 Grenoble, France

* Corresponding author: nicolas.bernier@cea.fr

# A. FIB sample preparation

In this study, both samples were prepared using a Ga+ focused ion beam (FIB) in a Zeiss Crossbeam 550 FIB-SEM microscope. To protect the regions of interest from ion beam damage, thick platinum protective layers were deposited. The lamellae were mounted on a copper TEM grid and initially thinned to approximately 700 nm using a 30 kV voltage and a 1–2 nA current. Subsequently, the specimens were refined by progressively lowering the voltage to 2 kV to minimize surface damage caused by the ion beam.

As EDS spectra are acquired using multiple detectors in the STEM, it is crucial to ensure that the solid angle for collecting x-rays is identical across all detectors. **Fig. S1a-b** illustrate this point. **Fig. S1a** shows a FIB-SEM image of an Omniprobe® lift-out grid finger, alongside a schematic representation of the x-ray collection geometry for two detectors. When the lamella is mounted on the side wall of a grid finger, shadowing effects can occur, where the finger itself obstructs the EDS detector. This issue is experimentally confirmed by measuring x-ray signals from each detector independently. Additionally, emitted x-rays can be absorbed by the thicker regions of the lamella, where metallic glue is deposited. As shown in **Fig. S1b,** positioning the thin portion of the lamella at the highest possible level ensures that detectors are not shadowed. To address these challenges, a dedicated FIB preparation protocol was developed.

First, the side wall of the finger beneath the lamella must be removed to avoid to block the transmitted electrons beam. This is achieved by rastering the Ga$^+$ beam over the area highlighted in yellow in **Fig. S1c**. The SEM image in **Fig. S1d** shows the finger after complete thinning of the side wall. Next, the finger is rotated 180° in the FIB, and the lamella is mounted on the front side of the finger, as shown in **Fig. S1e**. The final step involves thinning the lamella with a Ga+ beam at low voltage and current, ensuring that the entire upper surface of the lamella (the side first impacted by the electron beam in the STEM) is flat. The metallic contact to the tip should also remain continuous, as illustrated in **Fig. S1f**. Importantly, the upper surface of the lamella (marked by the green dotted line) is positioned above the highest level of the supporting grid (indicated by the orange dotted line), optimizing the configuration for EDS analysis.

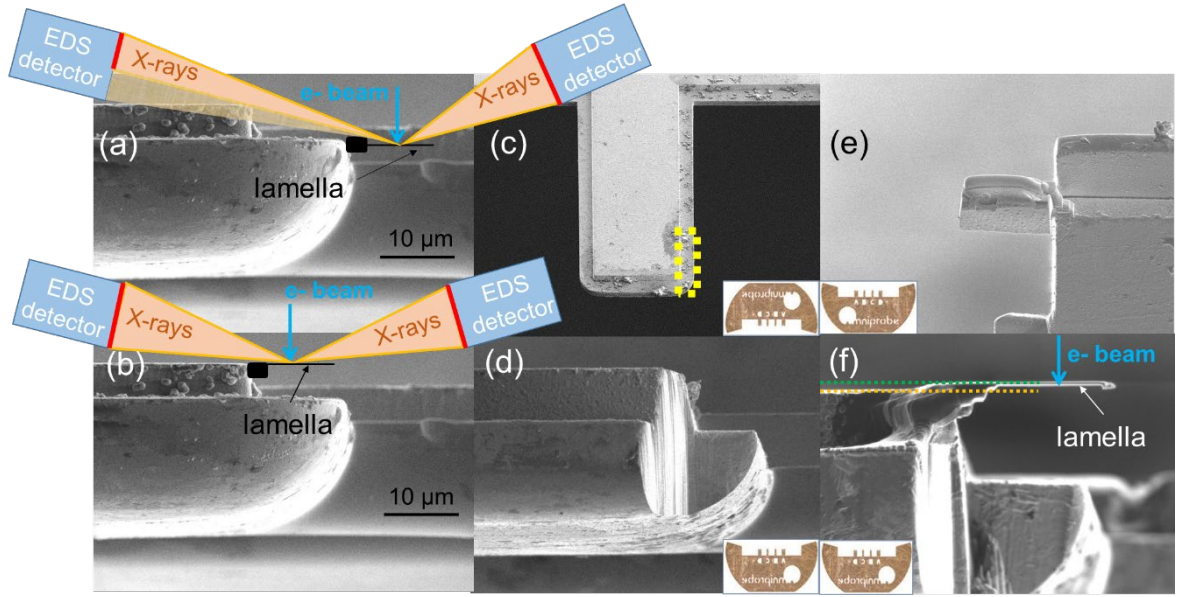

**Fig. S1. Sample preparation method optimized for quantitative EDX analyses.** (a-b) Schematic representation of how the position and shape of the lamella influence the shadowing of EDS detectors. (c-f) Step-by-step illustration of the optimized FIB preparation process to ensure identical solid angles for x-ray collection across all detectors: (c) FIB image of the Omniprobe® lift-out grid showing the FIB pattern (yellow dotted rectangle) used to remove the side wall of the finger beneath the lamella, (d) SEM image of the finger after the side wall has been completely removed, (e) FIB image of the lamella mounted on the upper front side of the finger, and (f) SEM image of the final thin lamella with a fully flattened upper surface (indicated by the green line) positioned above the highest level of the supporting grid (indicated by the orange line).

To determine the absolute sample thickness, low-loss STEM-EELS measurements were performed. **Fig. S2a** shows the HAADF image of the TQW sample, while **Fig. S2b** displays the corresponding relative thickness map ($t/\lambda$), derived from the ratio between the elastic and inelastic signals according to the log-ratio method:

$$\frac{t}{\lambda} = \ln \frac{I_t}{I_0}$$

where $\lambda$ is the inelastic mean free path (IMFP) of electrons in the material, and $I_0$ and $I_t$ are the elastic (zero-loss peak) and inelastic signals extracted from the EELS spectrum, respectively. A relative thickness profile extracted from this map is shown in **Fig. S2c**, to accurately locate the region of interest for HR-EDX analysis. The corresponding EELS data for the TQW heterostructure are presented in **Fig. S2d-f**, where the FIB lamella contains three windows of varying thickness. The regions used for HR-EDX thickness comparison (discussed in **section 4.3**) are indicated with red and green squares and arrows.

To convert the relative thickness to an absolute value, the estimation of the IMFP $\lambda$ is required. For the material studied here, the most reliable approach is the model proposed by Iakoubovskii *et al.* [1], which estimates $\lambda$ based on material density $\rho$, offering better accuracy than older models (e.g., Malis *et al.* [2]) that depend on mean atomic number:

$$\frac{1}{\lambda} = \frac{11\rho^{0.3}}{200FE_0} \ln\left[\frac{\alpha^2 + \beta^2 + 2\theta_E^2 + |\alpha^2 - \beta^2|}{\alpha^2 + \beta^2 + 2\theta_C^2 + |\alpha^2 - \beta^2|} \cdot \frac{\theta_C^2}{\theta_E^2}\right]$$

$$F = \frac{(1 + E_0/1022)}{(1 + E_0/511)^2}, \qquad \theta_E = \frac{5.5\rho^{0.3}}{FE_0}, \qquad \theta_C = 20 \text{ mrad}$$

where $E_0$ is the beam energy (in keV), and $\alpha$ and $\beta$ the convergence and collection angles, respectively.

In the heterostructure investigated, the local density ranges from $\rho = 6.15 \text{ g.cm}^{-3}$ (pure GaN) to $\rho = 5.88 \text{ g.cm}^{-3}$ (AlGaN layers with 5 at.% Al). The IMFP was calculated to be $\lambda = 120.1$ nm for pure GaN, accounting for the limited collection angle used during low-loss spectrum acquisition. This value increases slightly to $\lambda = 121.4$ nm in Al-rich regions. Consequently, the absolute thicknesses of the regions analyzed by HR-EDX are estimated to be approximately 38 nm in the thinnest area and 133 nm in the thickest region used in **section 4.2**.

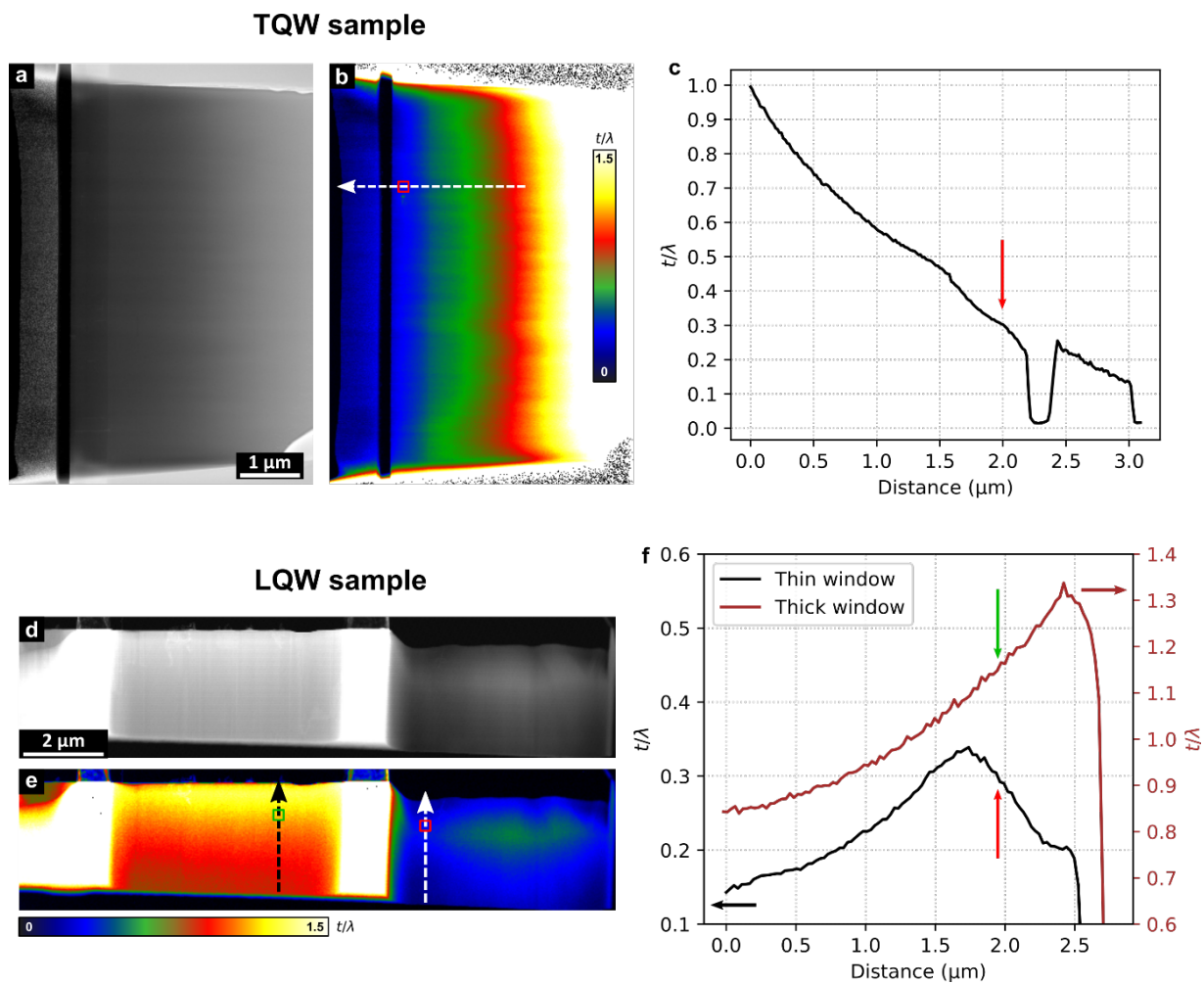

**Fig. S2**. Presentation of the samples and thickness mapping. (a) STEM-HAADF image of the TQW heterostructure, and (b) corresponding $t/\lambda$ map extracted from low-loss STEM-EELS hyperspectral image. (c) Relative thickness profile taken along the direction indicated by the white arrow in (b). The region used for HR-EDX acquisition is marked with a red square in (b) and a red arrow in (c). (d-f) Same analysis for the LQW sample, where three windows with different thicknesses were created. Thickness profiles are shown for the medium and thin windows.

## B. Beam propagation in pure GaN crystal

To eliminate the spurious Al K x-ray signal, we calibrated this radiation using the 10 nm-wide GaN layers present in the Large Quantum Well (LQW) sample. It was first necessary to confirm that, under these experimental conditions, the lateral beam propagation was not extensive enough to reach the surrounding AlGaN layers, as occurs in the Thin Quantum Well (TQW) sample. To validate this, simulations were performed to observe the beam propagation profile within pure GaN material and measure the extent of lateral beam diffusion as it traversed the sample. Three beam positions were analyzed, focusing on three different atomic columns: gallium (**Fig. S3a**), nitrogen (**Fig. S3b**), and an inter-position located between the two (**Fig. S3c**). Each graph illustrates the evolution of the probe's integrated intensity as a function of sample thickness, considering various integration factors ranging from 10% to 90% of the total beam intensity in the simulation grid. These representations provide direct insight into the lateral diffusion amplitude based on the beam's position, highlighting the pronounced impact of gallium columns on beam propagation. The red curves correspond to the commonly used 60% and 90% intensity thresholds to characterize the beam diameter [3], with the respective numerical values presented in **Tab. 1** of the main text. The graph also shows the 90% intensity beam diameter as determined by the analytical model of Goldstein *et al.* [4,5] a commonly used criterion to evaluate beam broadening in STEM. The results indicate that this simple analytical approach predicts beam broadening values approximately midway between those observed at the three different scan positions, providing a good overall approximation of the beam broadening across the entire unit cell. However, the simulations are essential to capture the variations between different atomic columns, which is necessary for a detailed understanding of the HR-EDX data. It is important to note that for thicknesses exceeding approximately 60 nm, some distortion due to wrap-around effects can be observed in the simulation for the Ga atomic column. This distortion explains why the external probe diameter (corresponding to the 90% beam fraction) could not be determined for a sample thickness of 133 nm.

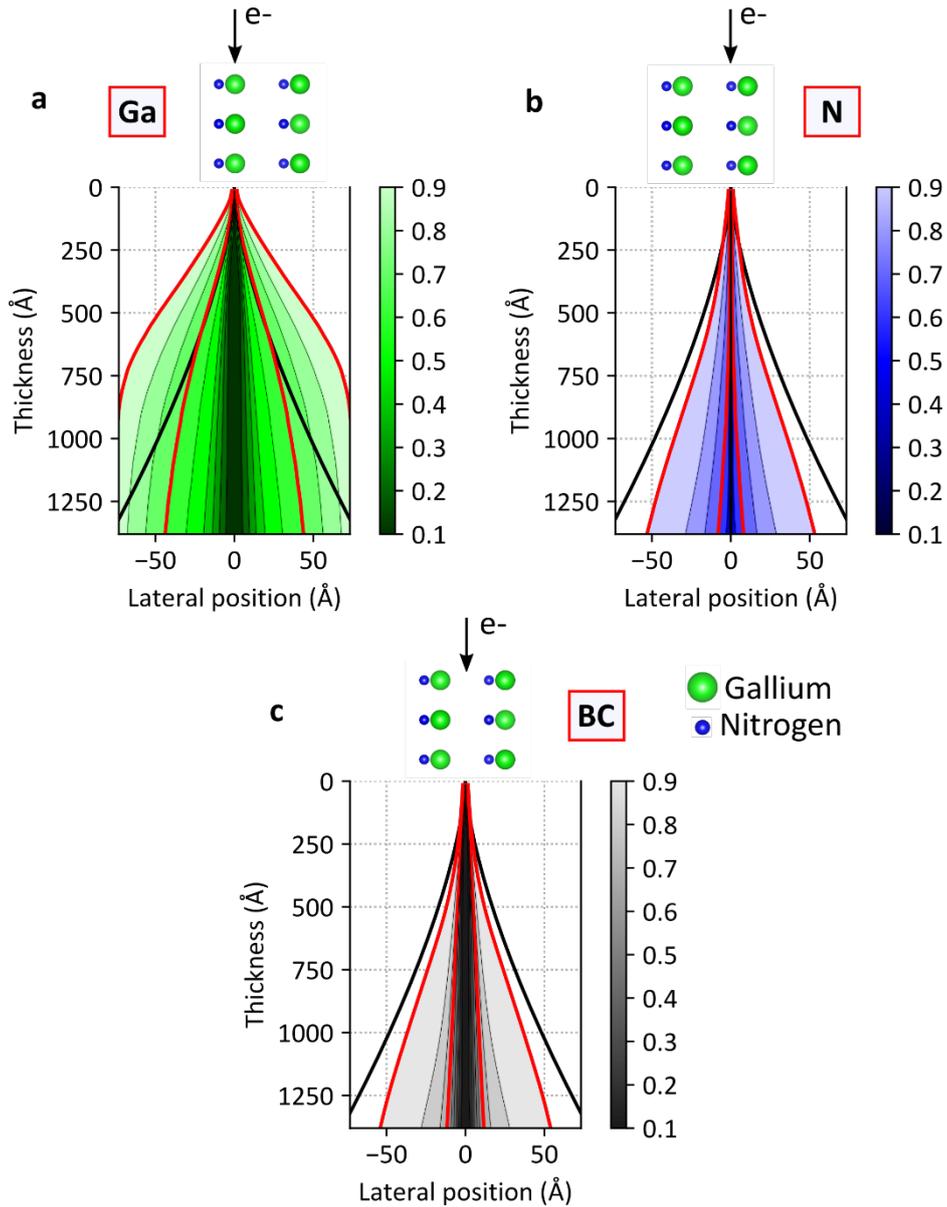

**Fig. S3. Evolution of the electron probe's integrated intensity with increasing GaN thickness** for three beam positions: (a) the gallium column position, (b) the nitrogen column position, and (c) the inter-column position. The color scale indicates the integration factor used to generate each iso-intensity contour. Crystal lattice diagrams, viewed perpendicular to the beam propagation direction, illustrate the three beam positions studied. Red contours highlight the 60% and 90% integrated intensity thresholds. The thick black lines represent the 90% intensity boundary ($b_{90}$), demonstrating that the single-scattering analytical model provides a good approximation of overall beam broadening behavior.

## C. Raw intensity maps

**Fig. S4** shows the raw intensity maps extracted from realigned STEM-EDX hyperspectral images acquired on the TQW sample. **Fig. S4a** displays the HAADF image, while **Fig. S4b-d** present the intensity distributions of the Al K, Ga K, and N K transitions, respectively. Despite the low signal-to-noise ratio (SNR), periodic features are clearly observable for all three chemical species, as confirmed by the Fourier transforms shown in the insets. These raw images were used to generate **Fig. 4** in the main text.

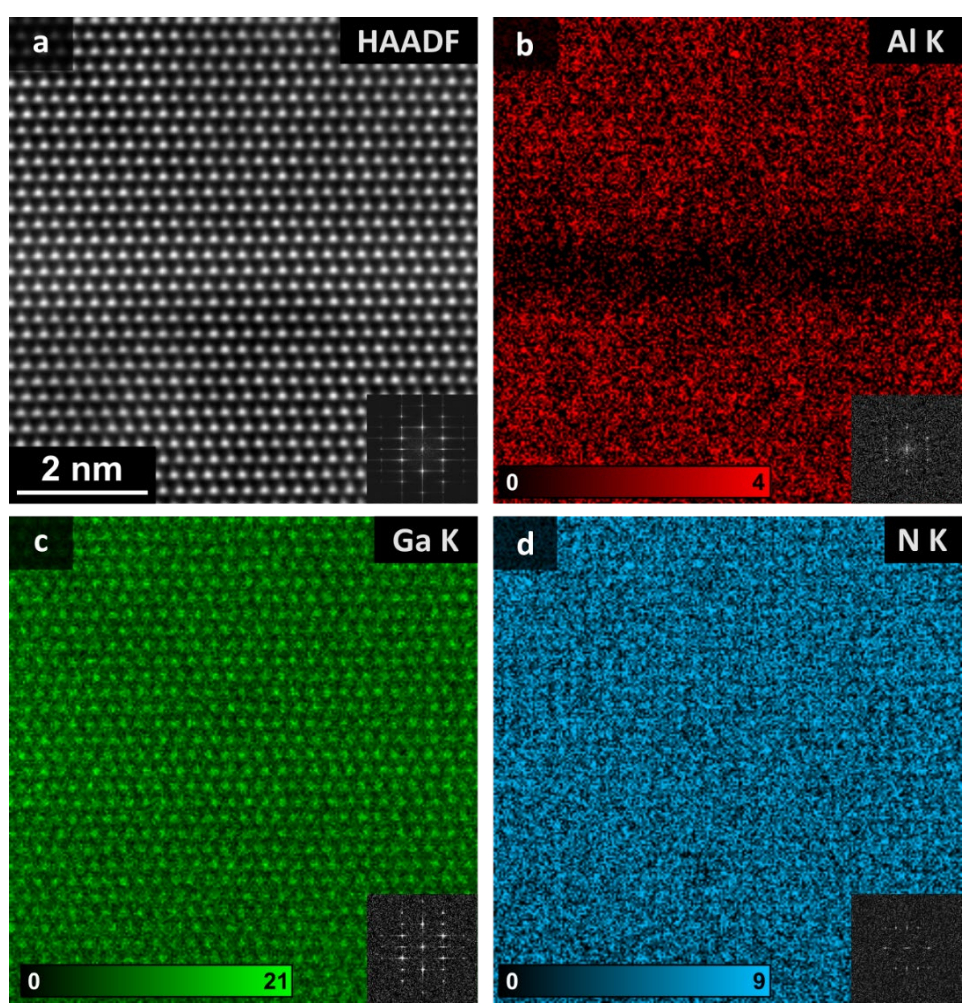

**Fig. S4.** (a) HAADF image and (b-d) raw elemental maps corresponding to the Al K, Ga K, and N K transitions. The insets present the Fourier transforms of the respective elemental maps.

# D. Calibration of ζ factors and calculation of the theoretical values

The determination of ζ-factors in STEM requires thin reference specimens with known composition and mass-thickness. Our approach is based on a fully experimental technique, which ensures consistency between SEM and STEM measurements by extending the conventional ζ-factor method. The key concept is the introduction of a modified factor, $\zeta^+$, which builds upon the standard ζ-factor method originally proposed by Watanabe and Williams [6], **Their approach assumes a constant x-ray generation rate along the beam path and applies simplified absorption corrections based on idealized geometries (e.g., infinite lamella). However, this conventional method is not well suited for low accelerating voltages or bulk specimens, where strong backscattering and absorption effects are common - especially in SEM.**

In our approach, the $\zeta^+$-factor incorporates two key correction terms, inherited from the ZAF model traditionally used in SEM-EDX quantification:

1. $\bar{\varphi}_i$ accounting for the x-ray generation depth profile $\varphi(\rho z)$ in the specimen;
2. $\bar{\chi}_i$ replacing the standard absorption coefficient to reflect the influence of the actual x-ray generation profile $\varphi(\rho z)$

**To generalize the model across experimental configurations, we also introduce a third correction** $\bar{\theta}$ corresponding to the compensation for the electron beam's incidence angle relative to the specimen. Within this framework, the new definition of the ζ-factor becomes:

$$C_i = q \frac{\zeta_i}{\rho z_\theta} \frac{I_i}{I_b} \frac{\bar{\chi}_i}{\bar{\varphi}_i}$$

**where** $z_\theta$ corresponds either to the physical thickness of a TEM lamella (for thin specimens) or to the effective ionization depth (in the case of bulk samples). On one side, $\bar{\theta}$ and $\bar{\varphi}_i$ are parameterized using Monte Carlo simulations performed with the CASINO software [7], covering a wide range of materials and experimental conditions. The absorption correction is adapted from the quadrilateral model developed by Scott and Love [8], using mass absorption coefficients from NIST [9].

Importantly, in the special case of a standard thin TEM sample studied under high voltage (100s keV), the model naturally reduces to the traditional ζ-factor framework (**Eq. 1** of the main text), assuming constant x-ray generation ($\bar{\varphi}_i = 1$) and purely geometric absorption coefficients ($\bar{\chi}_i = A_i$).

In practical terms, this methodology was applied for ζ-factor calibration in STEM using the following three-step procedure:

1. Calibration of ζ factors in an SEM at 30 kV using highly pure bulk reference samples, where the mass-thickness value is not required since the incident electron beam is fully stopped within the bulk material.

2. Fabrication of TEM lamellae from these same reference samples, followed by the measurement of mass-thickness ($\rho t$) in the SEM at 30 keV (same conditions that step 1) for each TEM foil using the previously calibrated ζ factors.

3. Measurement of ζ factors in the STEM at 200 kV using the well-characterized TEM lamellae from the reference samples.

For the elements of interest in this study, thin lamellae were prepared from pure AlN and GaN bulk specimens using a TFS Strata 400S dual-beam machine, following the same procedure described in **section A**. After determining the mass-thickness in the SEM, the thin samples were analyzed in STEM-EDX using the TFS Titan Themis microscope operated at 200 kV (same microscope used for the HR-EDX analyses) to extract the ζ factors specific to this instrument. The results are presented in **Tab. 2** of the main text. Uncertainties were determined through error propagation stemming from the mass-thickness determination in the SEM, which is the primary limiting factor.

The experimentally measured ζ factors were compared to theoretical values calculated using the following equation from Watanabe and Williams [10]:

$$\zeta_i = \frac{M_X}{N_v Q_i \omega_i a_i \varepsilon_i \varepsilon_c} \qquad 11$$

where $M_X$ is the atomic mass of element $X$, $Q_i$ is the total x-ray emission cross-sections at 200 kV for atomic transition $i$, $\omega_i$ is the fluorescence yield, $a_i$ is the relative intensities of the x-ray lines, and $\varepsilon_i$ is the detector efficiency for the transition considered. $N_v$ is the Avogadro's number and $\varepsilon_c$ is the collection efficiency ($\varepsilon_c = \Omega/4\pi$ with $\Omega$ the detector solid angle). Using the net x-ray intensities of both Kα and Kβ transitions for the three elements of interest, ζ factors were determined with $\alpha_i = 1$. The total emission cross-sections ($Q.\omega$) were provided from the NIST database [11], and evaluated at 83.0, 102.2 and 45.2 barns for the Al K, Ga K and N K lines, respectively. Considering the detectors properties (contact material, dead and active layers thicknesses), the detection efficiencies for these transitions were evaluated as 99.4%, 98.2% and 82.3%, respectively [12].

# E. Impact of AlGaN barriers content on the determination of quantum well composition

In the main text, a comparison is made between simulation results obtained from different chemical models of the AlGaN/GaN/AlGaN system. Since these models primarily explore variations in composition within the quantum well and at the interfaces, it became essential to investigate the potential effect of the chemical composition of the barriers included in the models. We recall that 5.3 ± 0.2 at.% of Al in the barriers is measured, in good agreement with the nominal values.

**Fig. S5** compares two models, both featuring the same compositional gradient at the interfaces and 0.25 at.% aluminum within the QW, but differing in the aluminum content of the AlGaN barriers: one with the nominal composition of 5.0 at.% Al and the other slightly richer, with 5.5 at.% Al. It is important to note that the Al K intensity signal in the barriers is normalized for each theoretical curve to allow comparison with the experimental data, resulting in identical normalized Al K intensity values in the barriers (**Fig. S5a**). The proper parameter analyzed is the modification of $\delta I_{Al} = I_{Al} / \overline{I_{Al}^{bar}}$, the relative intensity of the Al K signal within the QW compared to the barriers. As shown in **Fig. S5b**, the variation in barrier composition has a negligible effect on $\delta I_{Al}$, much smaller than the impact of changes in the composition within the QW or at the interfaces explored in the main text. This result confirms the reliability and robustness of the proposed approach.

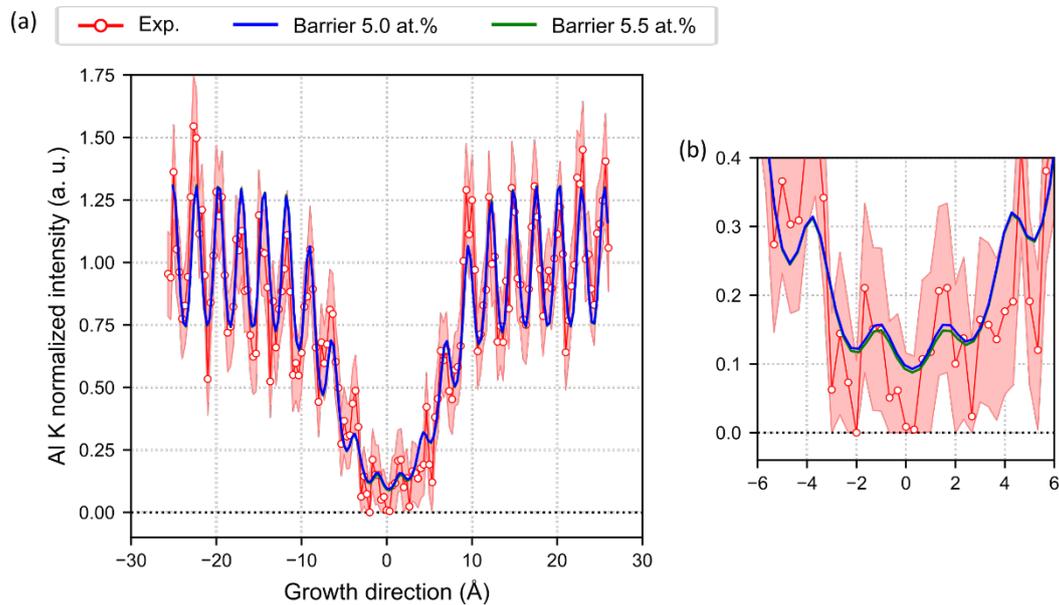

**Fig. S5**. **Comparison of experimental and simulated Al K intensity profiles** based on two different chemical models of the heterostructures, varying the aluminum content in the barriers. (a) Experimental Al K intensity profile compared with the results of the two simulations, and (b) a detailed view focusing on the quantum well region.

## F. Channeling effects in nitrides

A specific issue arises from the requirement to tilt samples along a zone axis to resolve the atomic columns of the material. In this orientation, the channeling effect significantly influences the inelastic cross-section of various scattering events [13]. This introduces a non-linear relationship between the measured x-ray intensity and the element concentration, potentially affecting quantitative results [14,15]. In the present case, the studied emission transitions exhibit a large energy difference (notably between the N K and Ga K peaks), resulting in differing intensity variations between on-axis and off-axis conditions, thereby altering the relative concentrations measured. The extent of such quantification deviations under varying experimental conditions must be assessed using a reference sample to determine optimal experimental parameters.

To address this, we investigated the variations in measured concentration within a pure GaN substrate, following the ζ-factor procedure described in the main text. GaN's stoichiometry makes it an ideal model for detecting potential concentration deviations. **Fig. S6** shows the relative composition deviation (in %) from the compound's stoichiometry (50 at.% for both species) as a function of diffraction conditions: along the zone axis (on-axis) and off the zone axis direction (off-axis), where the specimen was tilted by more than 100 mrad. The comparison was conducted for different convergence angles $\alpha$, as previously reported to be a key factor influencing channeling effects [16].

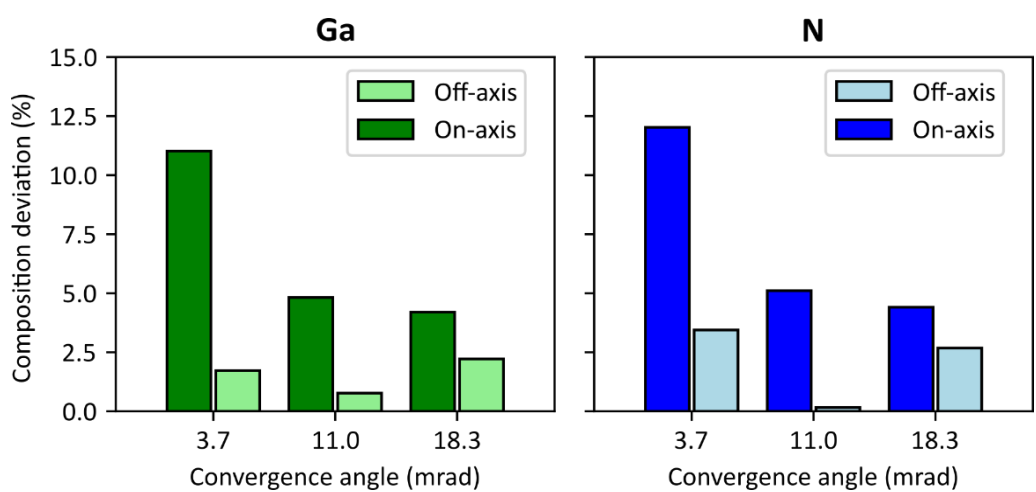

**Fig. S6**. **Composition deviation measured on a stoichiometric GaN substrate** as a function of convergence angle and diffraction conditions.

When the crystal is tilted away from dynamic diffraction conditions (off-axis), deviations are significantly reduced, averaging less than 2% relative deviation from the nominal concentration. Conversely, results obtained along the zone axis reveal deviations significantly different from the reference composition. Quantitative measurements deviate most from the reference value at low convergence angles, where channeling effects are stronger, but these deviations diminish rapidly as $\alpha$ increases. Beyond 11 mrad, the relative deviation for both chemical elements approaches 4%, slightly higher than the deviations measured off-axis.

These results indicate that while channeling effects lead to a substantial increase in x-ray production, their impact on the measured chemical composition of nitrides is relatively restricted compared to the uncertainties associated with x-ray intensities, as demonstrated in the main text.